\documentclass[a4paper,fleqn,usenatbib]{mnras}
\usepackage[T1]{fontenc}
\usepackage{ae,aecompl}

\usepackage{graphicx} 
\usepackage{amsmath}  
\usepackage{amssymb}  
\usepackage{bm}
\usepackage{subfig}
\usepackage{mathtools}
\usepackage{float}
\usepackage{times}
\usepackage{color}
\usepackage{amsfonts}
\usepackage{booktabs}
\usepackage{siunitx}
\usepackage{soul}
\usepackage{xcolor}
\usepackage{amsmath}
\usepackage{float}
\usepackage{capt-of}
\usepackage{multicol}
\usepackage{multirow,multicol}

\newcommand{\eg}{e.g.,~}
\newcommand{\ie}{i.e.,~}
\newcommand{\cf}{cf.,~}

\title[Jet-ejecta interactions]{Impact of anisotropic ejecta on jet
  dynamics and afterglow emission in binary neutron-star mergers}

\author[V. Mpisketzis et al.] {Vasilis Mpisketzis$^{1,2}$, Rapha\"el
  Duqu\'e$^{1}$, Antonios Nathanail$^{3}$, Alejandro Cruz-Osorio$^{6,1}$
  \newauthor and Luciano Rezzolla$^{1,4,5}$ \\
  $^{1}$Institut f\"ur Theoretische Physik, Goethe Universit\"at
  Frankfurt, Max-von-Laue-Str.1, 60438 Frankfurt am Main, Germany \\
  $^{2}$Department of Physics, National and Kapodistrian University of
  Athens, Panepistimiopolis, GR 15783 Zografos, Greece \\
  $^{3}$Research Center for Astronomy, Academy of Athens, Soranou 
  Efessiou 4, 115 27 Athens, Greece  \\
  $^{4}$School of Mathematics, Trinity College, Dublin 2, Ireland\\
  $^{5}$Frankfurt Institute for Advanced Studies, Ruth-Moufang-Str. 1,
  60438 Frankfurt am Main, Germany\\
  $^{6}$Instituto de Astronom\'{\i}a, Universidad Nacional Aut\'onoma de M\'exico, AP 70-264, Ciudad de M\'exico 04510, M\'exico}

\usepackage{hyperref}
\hypersetup{
  colorlinks=true,        
  linkcolor=blue,         
  citecolor=cyan,         
  filecolor=magenta,      
  urlcolor=magenta        
}

\begin{document}
\label{firstpage}
\pagerange{\pageref{firstpage}--\pageref{lastpage}}
\maketitle

\begin{abstract}
Binary neutron stars mergers widely accepted as potential progenitors of
short gamma-ray bursts. After the remnant of the merger has collapsed to
a black hole, a jet is powered and may breakout from the the matter
expelled during the collision and the subsequent wind emission. The
interaction of the jet with the ejecta may affect its dynamics and the
resulting electromagnetic counterparts. We here examine how an
inhomogeneous and anisotropic distribution of ejecta affects such
dynamics, dictating the properties of the jet-ejecta cocoon and of the
afterglow radiated by the jet upon deceleration. More specifically, we
carry out general-relativistic hydrodynamical simulations of relativistic
jets launched within a variety of geometrically inhomogeneous and
anisotropic distributions of ejected matter. We find that different
anisotropies impact the variance of the afterglow light-curves as a
function of the jet luminosity and ejected mass. A considerable amount of
the jet energy is deposited in the cocoon through the jet-ejecta
interaction with a small but important dependence on the properties of
the ejecta. Furthermore, all configurations show a two-component
behaviour for the polar structure of the jet, with a narrow core at large
energies and Lorentz factors and a shallow segment at high latitudes from
the jet axis. Hence, afterglows measured on off-axis lines of sight could
be used to deduce the properties of the ejected matter, but also that the
latter need to be properly accounted for when modelling the afterglow
signal and the jet-launching mechanisms.
\end{abstract}


\begin{keywords}gamma-ray burst: general, hydrodynamics, relativistic processes, stars:neutron 
\end{keywords}

\section{Introduction} 
\label{sec:intro}

The association between binary neutron-star (BNS) mergers and short
gamma-ray bursts (SGRBs) was recently consolidated in August 2017 via the
coincident detection of the gravitational-wave signal from the inspiral
of a BNS (GW170817, \citealt{Abbott2017}) and of a short, hard gamma-ray
burst (GRB170817, \citealt{Goldstein2017,Abbott2017d}). Although within
the population of SGRBs observed so far this event is characterised by a
very low luminosity~\citep{2019MNRAS.483.1247M}, its multi-messenger
nature provides strong evidence in support of the interpretation of BNS
mergers as progenitors of SGRBS \citep{1986ApJ...308L..43P,Eichler89}. In
the case of GW178717, in particular, the afterglow observations suggest
that a regular, bright SGRB would have been observed if it has been seen
from an on-axis line of sight~\citep{2019A&A...628A..18S}.

After these signals were first detected, a follow up campaign allowed to
localise the source via the detection of the kilonova optical transient
(\citealt{2017ApJ...848L..12A} and references therein). Spectroscopic
observations of this transient suggests the presence of $r$-process
nucleosynthesis in the merger ejecta \citep{2017Natur.551...75S,
  2019Natur.574..497W}. The discovery and long-term sampling of the
afterglow signal across the electromagnetic spectrum opened a rich window
on the outflow from the merger \citep{2017Natur.551...71T,
  2017Sci...358.1579H, 2018NatAs...2..751L}. The combined photometry and
VLBI image of the afterglow gave clear indication of a relativistic jet
emerging from the source, being viewed with a significantly off-axis
line-of-sight \citep{Mooley2018,Ghirlanda2019}. The multi-wavelength
light-curves of GW170817 were characterized by a late-time power-law
decay. In addition, observations using very long baseline interferometry
revealed an apparent superluminal motion \citep{Mooley2018, 2019ApJ...870L..15L,
  Ghirlanda2019}. This motion helped constrain the range of allowed
observational angles to $\theta_{\mathrm{obs}} = 15\pm 1.5
\ \mathrm{deg}$ at 1-$\sigma$ and hinted to a very energetic jet with a
core energy of $E_{\rm jet} \approx 10^{49-50}\, \mathrm{erg}$; other
works \citep{2021MNRAS.502.1843N,2023ApJ...943...13W} extended the
inclination range to even higher values, and close to $30 \,
\mathrm{deg}$. The quantity of ejecta, inferred from the kilonova
transient suggests that the remnant object collapsed at a time of
$\approx 1 \mathrm{sec}$ after the merger \citep{2019ApJ...876..139G,
  murguia-berthier2020, 2020ApJ...895L..33B}. Immediately afterwards, the
energetic jet was launched within the massive envelope, which was moving
with semi-relativistic velocities~\citep[see, \eg][]{rezzolla:2011,
  2013PhRvD..87b4001H, Palenzuela2015, Ruiz2016}.

Considerable theoretical work has been invested over the years to study
the evolution of SGRB jets, starting from the launching of the incipient
jet in the immediate environment of the mergers \citep[see,
  \eg][]{2005A&A...436..273A, rezzolla:2011, murguia-berthier2016,
  2019ApJ...884...40M, Hayashi2021a, 2021MNRAS.506.3511M}, continuing
with its interaction with the ejecta and breakout
\citep{2021ApJ...915L...4G, 2021MNRAS.503.4363U, 2022MNRAS.517.1640G},
and up to the exploration of the various processes that shape the
large-scale structure of the jet and that eventually lead to the
afterglow signal. Among these processes, the interaction with the merger
ejecta is expected to be particularly interesting as the jet can
potentially acquire (or modify) its structure, hence determining how the
Lorentz factor and energy of the material in the jet at a large distance
from the source depend on the latitude with respect to the jet
axis. These dependencies on the latitude in the jet are usually captured
in two functions $\Gamma_{\infty}(\theta)$ and ${\rm d}E_{\rm jet}/{\rm
  d}\Omega(\theta)$ denoting the Lorentz factor and distribution of total
energy of the jet material per unit solid angle at latitude $\theta$ in
the jet, which is assumed to be axis-symmetric.  Together, these
functions as referred to as the ``jet structure''~\citep[see, e.g.,][for
  a review]{2022arXiv220611088S}.

Hence, understanding the formation of the jet structure in this chain of
events is not only interesting, but also important when making use of the
afterglow emission to infer the structure of the relativistic jet and, in
turn, to explore the lower-scale physics of jet launching.

This connection between the large-scale structure of the jet and the
afterglow signal has been explored a number of times in the literature
and employing a variety of methods: 
\begin{itemize}
  \item by prescribing a functional form for the structure and by
    inferring its parameters through light-curve fitting \citep[see,
      \eg][for the case of
      GW170817]{gill2018,2019MNRAS.489.1919T};
  \item by deriving a standard inversion procedure to determine the
    structure starting from the light-curve
    \citep[see, \eg][]{2021MNRAS.501.5746T};
  \item by discovering a set of jet structures all consistent with the
    data from GW170817 \citep[see, \eg][]{2020MNRAS.497.1217T};
  \item by studying analytically the diverse light-curve morphologies
    that can arise from a given form of the jet structure \citep[see,
      \eg][]{2020MNRAS.493.3521B};
  \item by focusing on the mimicking of jet-structure effects by other
    jet dynamics, such as its lateral expansion \citep[see,
      \eg][]{2021MNRAS.506.4163L}.
\end{itemize}
What all of these methods have in common is the realisation that
degeneracies exist in the theoretical modelling and hence all the
available information needs to be incorporated when constraining the jet
structure from afterglow observations.

The connection between the BNS merger and the large-scale structure of
the jet is tightly linked with the dynamical evolution that takes place
in the post-merger. More specifically, the violent merger produces an
immediate dynamical ejection of mass \citep[see, \eg][]{Rezzolla_2010,
  2015PhRvD..91f4059S, 2016CQGra..33r4002L, Radice_2018,
  2020PhRvD.102l4077R}, accompanied by secular mechanisms that further
eject mass from the system through magnetically-driven or and
neutrino-driven winds from the accretion disk and from the remnant before
its eventual collapse \citep{2009ApJ...690.1681D,siegel2014,
  2015MNRAS.446..750F,2016PhRvD..94l3016F,2018ApJ...860...64F}.
Subsequently, and on much larger scales away from the central object, the
jet-ejecta interaction leads to the formation of a jet-cocoon
system~\citep{2011ApJ...740..100B} and imprints the jet structure upon
breakout, before it propagates to larger distances and produces gamma-ray
burst prompt and afterglow radiation. General-relativistic simulations
have been instrumental in the study of this early phase of the jet
dynamics and has allowed, among other things, to study the conditions for
the jet to breakout~\citep{Duffell2018,2022ApJ...933L...2G}, the effect
of baryon mixing between the jet and the cocoon~\citep{
  Gottlieb2020b,Gottlieb2020}, the possible formation of a hollow-core
structure and its dependence on magnetic
fields~\citep{2019ApJ...870L..20N,nathanail2020b}, the role of the
structure of the incipient jet in its interaction with the
ejecta~\citep{2021MNRAS.503.4363U}, or the possible asymmetries and
oscillations of the jet imprinted by the interaction with the ejecta due
to three-dimensional~\citep{2021ApJ...918L...6L}. Furthermore, by
performing simulations of the jet-ejecta interaction and comparing
spherically-symmetric ejecta with more realistic ejecta imported from BNS
merger simulations,~\citet{pavan2021a} have pointed out the considerable
impact that even a small anisotropy in the ejecta rest-mass density
profile can have on the emerging jet structure. These numerical studies
use various prescriptions for the ejecta and incipient jet properties,
and include various sets of physical processes in the simulations.

In this work, we extend the exploration of the impact of the jet-ejecta
interactions on the large-scale properties of the jet to the case of an
\textit{inhomogeneous} and \textit{anisotropic} distribution of matter in
the merger ejecta. These radial and angular gradients can be produced by
a number of processes, such as, unequal composition and heating rates by
nucleosynthetic processes stemming from anisotropic neutrino
irradiation~\citep{2023ApJ...944...28C}, or by sporadic baryon loading of
the base of the jet, to name a few~\citep{dionysopoulou2015,
  2019MNRAS.488.1416G}. Furthermore, the different components of the
merger ejecta (dynamical ejecta, winds, etc.) are expected to mix and
form shocks \citep{Bovard2017, Radice_2018}, resulting in an overall
inhomogeneous envelope which the jet will have to penetrate.

In particular, we carry out two-dimensional (2D) general-relativistic
hydrodynamic simulations of relativistic jets propagating through an
envelope of ejected material following a variety of prescribed rest-mass
density profiles featuring strong gradients in both the radial and polar
direction. We study the interaction of the jets for up to $3$\,s after
jet launching, focusing on the effects of these gradients on the
jet-cocoon system, on the propagation of the jet and on the large-scale
structure adopted by the jet after breakout. Overall, we find that the
cocoon is generally robust to different anisotropies in the ejecta and
that the more massive the ejecta are relative to the jet energy, the more
the jet structure is affected. As a result, a strong observational
imprint is in principle present on off-axis afterglow observations of the
most massive ejecta.

The structure of the paper is organised as follows. In
Sec.~\ref{sec:setup} we describe the numerical methods employed for both
the numerical simulations and the calculation of the afterglow,
accompanied by the different setups for the ejected matter and the
injection of the jet. We instead report the results of the simulations
and afterglow observations in Sec.~\ref{sec:results}, together with
comparison with the observations and how they can be used to infer the
properties of the ejected material. The conclusions and prospects for
future work are presented in Sec.~\ref{sec:conclusion}.

\section{Methods} 
\label{sec:setup}

\subsection{Numerical setup}
\label{sec:bhac}

Our simulations are performed employing the numerical code
\texttt{BHAC}~\citep[][]{Porth2017,Olivares2019}, which solves the
equations of general-relativistic hydrodynamics or magnetohydrodynamics
using high-resolution, shock capturing methods in one, two and three
spatial dimensions and in a variety of coordinates (we have employed
spherical polar coordinates here). Adopting the Schwarzschild solution as
a fixed curved background metric, our grid starts from an inner radius of
$R_{\mathrm{in}} = 10^{9}\,\mathrm{cm}$ and extends to $R_{\mathrm{out}}
= 6 \times 10^{10}\,\mathrm{cm}$. While for the outer boundary we impose
standard outgoing boundary conditions, a more involved prescription is
used for the inner ones as long as the jet is injected (see discussion
below for when the jet is progressively shut-down). More specifically, on
the first radial shell we examine whether the polar angle $\theta$ is
smaller than the opening angle of our jet $\theta_\mathrm{jet}$. If so,
we set an initial Lorentz factor and fix density and pressure in order to
match a top-hat jet, for a given luminosity $L_{\mathrm{jet}}$, and
asymptotic Lorentz factor $\Gamma_{\infty}$. For larger angles, we set a
smooth solution that follows the prescriptions detailed below for the
ejected matter, that extends from the inner boundary radius
$r_{\mathrm{in}} = R_{\mathrm{in}} =\ 10^{9}\,\mathrm{cm}$ out to
$r_{\mathrm{out}} = 5 \times 10^{10}\,\mathrm{cm}$.

\begin{figure*}
\centering \includegraphics[width=0.9\textwidth]{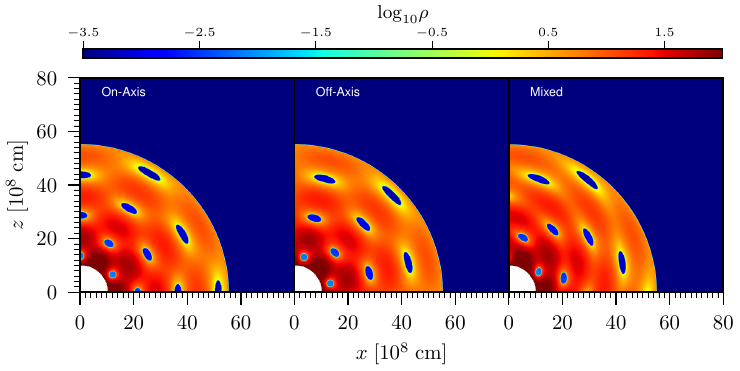}
\caption{Initial rest-mass density distributions of the ejecta considered
  in our set of simulations. Note that the distributions are both
  inhomogeneous and anisotropic, with voids that are either ``on-axis''
  (left panel), `` off-axis'' (middle panel), or ``mixed'' (right panel)
  [see Eq.~\eqref{eq:rho_rth} for details].}
\label{init}
\end{figure*}

The plasma is modelled as a simple ideal-gas
fluid~\citep{Rezzolla_book:2013} with adiabatic index $\gamma = 4/3$ and,
as anticipated above, is characterised by a large degree of
  inhomogeneity and anisotropy, as expected for the matter ejected on
  dynamical and secular timescales. Because the actual degree depends
  sensitively on the mass ratio of the binary, the equation of state and
  the magnetisation, it is difficult to derive from first principles
  analytic expressions that capture this large diversity. Of course, one
  could resort to importing the distribution of the ejecta directly from
  numerical simulations, but this would limit significantly the scenarios
  considered to those for which numerical simulations are performed and
  essentially prevent the systematic exploration that wish to carry out
  in our work. In view of these considerations, we have opted for a
  simplified prescription of the inhomogeneities and anisotropies in the
  ejected matter in terms of a reference ``homogeneous'' and isotropic
configuration in which the rest-mass density follows a power-law fall in
radius (see, e.g., \citealt{2014ApJ...784L..28N})
\begin{equation}
\rho(r) = \rho_0 \bigg(\dfrac{r}{r_{\mathrm{out}}}\bigg)^{-2}\,,
\label{ej:den}
\end{equation}
where $\rho_0$ is a constant. On the other hand, to study the impact of
the inhomogeneous and anisotropic ejecta on the dynamics and emission of
the jet, the rest-mass distribution is taken to have also a polar
dependence of the type 
\begin{equation}
  \label{eq:rho_rth}
  \rho(r,\theta) = \xi(r, \theta) \, \rho_0 \bigg(\dfrac{r}{r_{\mathrm{out}}}\bigg)^{-2}\,,
\end{equation}
where we vary the dimensionless function $\xi$ to explore three different
anisotropic configurations. More precisely, we further define
\begin{equation}
\xi(r, \theta) := \left\{ \begin{array}{ll}
1 - \phi (r,  \theta) & {\rm if}\, \phi (r,  \theta) < 0.85\,; \\ & \\
10^{-4} & {\rm else}\,.
\end{array} \right.
\end{equation}
and ensure that low-density patches exist in the ejecta by setting the
function $\phi(r,\theta)$ to:
\begin{equation}
  \label{eq:phi_rth}
  \phi = \left\{
  \begin{array}{l}
  \!\!\!\sin{(6\theta)}\sin{(6\pi(r\!-\!r_{\mathrm{in}})/r_{\mathrm{out}})}\,;\\ \\ 
  \!\!\!\cos{(6\theta)}\sin{(6\pi(r\!-\!r_{\mathrm{in}})/r_{\mathrm{out}})}\,;\\ \\ 
  \!\!\!\cos{(6\theta+ 2\pi(r\!-\!r_{\mathrm{in}})/(r_{\mathrm{out}}-
    r_{\mathrm{in}}))}\sin{(6\pi(r\!-\!r_{\mathrm{in}})/r_{\mathrm{out}})}\,.\\
  \end{array}
  \right.
\end{equation}
which we refer to as: ``on-axis'', ``off-axis'', and ``mixed'',
respectively; clearly, the ``homogeneous'' profile~\eqref{eq:rho_rth}
corresponds to the case in which $\phi(r,\theta)=0$. We note that while
the profiles given by the expressions~\eqref{eq:phi_rth} are simplified,
their spectral distribution in wavelengths resembles that observed in
numerical simulations; in addition, the ability to place in precise
locations regions of over- or under-density allows us to ascertain in
much more controlled way the impact that inhomogeneities have on the jet
propagation and hence on the afterglow signal (see discussion in
Sec.~\ref{sec:agres}). The initial conditions in the ejecta in the three
cases on-axis, off-axis or mixed can be found in the three panels of
Fig.~\ref{init}.

In all of the cases considered, the constant $\rho_0$ is determine such
that the total rest-mass in the ejecta, denoted by $M_{\rm ej}$, varies
in the range $M_{\mathrm{ej}}/M_\odot \in [0.001,\, 0.01]$, and
corresponds to the range expected from BNS mergers \citep{Rezzolla_2010,
  rezzolla:2011, 2013ApJ...773...78B, 2013PhRvD..87b4001H,
  10.1093/mnras/stw1227, 2017PhRvD..95d4045D, Radice_2018,
  2023ApJ...942...39F}.

For a given choice of ejecta distribution and total rest-mass $M_{\rm
  ej}$, the jet will sweep on its path is a rest-mass column defined as
\begin{equation}
M_{\mathrm{path}} := \int_{R_{\mathrm{in}}}^{R_{\rm{out}}}
\int_{0}^{2\pi} \int_{0}^{\theta _{\mathrm{jet,i}}} \rho(r,\theta) r^2 \,
\sin\theta \,d\theta\,d\phi dr\,,
\label{jetmass}
\end{equation}
which does not change by more than $5\%$ across the various profiles and
is at least $0.4\%$ of $M_{\mathrm{ej}}$.

By construction, the matter in the ejecta is all gravitationally unbound
and the corresponding initial velocity profile is assumed to be isotropic
and purely radial, increasing linearly with radius, \ie
\begin{equation}
  \label{eq:uu}
  \boldsymbol{v}_{\mathrm{ej}} = 0.3
  \left(\dfrac{r}{r_{\mathrm{out}}}\right)\boldsymbol{e}_r\,,
\end{equation}
where $\boldsymbol{e}_r$ is the unit-vector in the radial
direction. Expression~\eqref{eq:uu} leads to values that are close to
those measured in numerical simulations~\citep[see, \eg][for some
  reviews]{Baiotti2016, Paschalidis2016, 2019ARNPS..69...41S}, and within
the ranges inferred from the observations~\citep[see,
  \eg][]{2017Natur.551...64A, 2017Sci...358.1570D, 2017Sci...358.1574S,
  2017Natur.551...67P, 2017Sci...358.1583K, 2017Sci...358.1559K,
  2017Natur.551...80K, 2017Sci...358.1565E, 2017Natur.551...75S}.

The jet is assumed to be launched by the ``central engine'' at the
innermost radial shell and with an initial opening angle
$\theta_\mathrm{jet,i} = 0.1$. Its structure is assumed to be that of a
``top-hat'' jet, that is, with sharp cut-offs for all quantities at
angles larger than the opening angle. For each simulation, we inject a
jet with constant luminosity, $L_{\mathrm{jet}}$, using three different
reference values $10^{49}\,\mathrm{erg/s}$, $5\times
10^{49}\,\mathrm{erg/s}$, and $10^{50}\,\mathrm{erg/s}$. The jet is
injected in the grid for a duration of $t_{\mathrm{jet}} = 1\,
\mathrm{s}$ and has an initial Lorentz factor $\Gamma_0 = 10$. We set the
asymptotic value of the Lorentz factor to $\Gamma_{\infty} = 100$, which
is translated to an initial specific enthalpy for the injected plasma of
$h_0 = {\Gamma_{\infty}}/{\Gamma _0}$ since $h_{\infty}=1$. In practice,
at every timestep and for the grid cells in the first radial slice within
the jet opening angle we set
\begin{eqnarray}
&&\Gamma = \Gamma_0\,,  \nonumber\\
&&\rho = \rho_{\mathrm{jet}} = 2\pi
  \left(\dfrac{L_{\mathrm{jet}}}{\rho^2h_0\Gamma_0^2}\right)
  \left(\dfrac{1}{1-\cos(\theta_{\mathrm{jet}})}\right)\,,
  \\ \nonumber
&&p = p_{\mathrm{jet}} = \dfrac{1}{4}\rho_{\mathrm{jet}}(h_0 - 1)\,. 
\end{eqnarray}

\begin{table*}
  \centering
  \setlength{\tabcolsep}{1pt}
  \renewcommand{\arraystretch}{1.5}
  \begin{tabular}{|l|c|c|c|c|c|c|c|c|c|}
    \hline
        {name} &
        \scriptsize{\texttt{Lj.1.49.Me.001}} &
        \scriptsize{\texttt{Lj.1.49.Me.004}} &
        \scriptsize{\texttt{Lj.1.49.Me.010}} &
        \scriptsize{\texttt{Lj.5.49.Me.001}} &
        \scriptsize{\texttt{Lj.5.49.Me.004}} &
        \scriptsize{\texttt{Lj.5.49.Me.010}} &
        \scriptsize{\texttt{Lj.1.50.Me.001}} &
        \scriptsize{\texttt{Lj.1.50.Me.004}} &
        \scriptsize{\texttt{Lj.1.50.Me.010}} \\
        \hline
            {$L_\mathrm{jet}$} &
            \multirow{2}{*}{$1 \times 10^{49}$} &
            \multirow{2}{*}{$1 \times 10^{49}$} &
            \multirow{2}{*}{$1 \times 10^{49}$} &
            \multirow{2}{*}{$5 \times 10^{49}$} &
            \multirow{2}{*}{$5 \times 10^{49}$} &
            \multirow{2}{*}{$5 \times 10^{49}$} &
            \multirow{2}{*}{$1 \times 10^{50}$} &
            \multirow{2}{*}{$1 \times 10^{50}$} &
            \multirow{2}{*}{$1 \times 10^{50}$} \\
                     {$[{\rm erg/s}]$} & & & & & & & & & \\
                     \hline
                         {$M_{\rm ej}$}&
                         \multirow{2}{*}{$0.001$} & \multirow{2}{*}{$0.004$} &
                         \multirow{2}{*}{$0.010$} & \multirow{2}{*}{$0.001$} &
                         \multirow{2}{*}{$0.004$} & \multirow{2}{*}{$0.010$} &
                         \multirow{2}{*}{$0.001$} & \multirow{2}{*}{$0.004$} &
                         \multirow{2}{*}{$0.010$} \\
                                  {$[M_{\odot}]$} & & & & & & & & & \\
                                  \hline
  \end{tabular}
  \bigskip
  \caption{Properties of the various simulations performed in our
    analysis, each of which is marked by the corresponding jet luminosity
    $L_{\mathrm{jet}}$ and by the ejected mass $M_{\mathrm{ej}}$.}
\label{tab:runs}
\end{table*}

After a time $t=t_{\mathrm{jet}}$, which can be smaller or larger than
the breakout time depending on its energy and mass of the ejecta, the
central engine is not turned-off abruptly but following an exponential
fall-off for a duration of $\tau_{\mathrm{rel}} = 0.1
\ t_{\mathrm{jet}}$, which, in turn, produces a sufficiently large decay
of the asymptotic Lorentz factor. This smooth turning-off of the jet is
implemented to avoid the generation of artificial shocks in the numerical
domain, especially in the tail of the jet. After the transition time, we
treat the inner radial boundary inside the jet-angle in the same way as
we do for the ejected material. In other words, we set the rest-mass
density to follow a power-law in time with exponent $-6$. A summary of
the physical properties of the 27 different simulations carried out in
our investigations can be found in Table \ref{tab:runs}.

Finally, our 2D grid has a number of base cells in the radial and polar
direction given by $N_r \times N_{\theta} = 1120 \times 576$ and we
employ three refinement levels, thus reaching a peak effective resolution
of $N_r \times N_{\theta} = 8960 \times 4608$. This is a rather high
resolution that provides us with sufficiently detailed information about
the dynamics of the jet during its propagation and
breakout. Furthermore, a discussion of the consistency and robustness of
our results when considering different numerical resolutions can be found
in Appendix~\ref{sec:conv}.

\subsection{Large-scale jet structure and afterglow emission}
\label{sec:agmeth}

In order to determine the astronomical observables that would allow us
to probe the jet-ejecta interaction, we calculate the afterglow radiation
produced as a result of the jet morphology computed in the various
simulations. To this scope, we first need to compute the angular
structure in total jet energy ${\rm d}E_{\rm jet}/{\rm d}\Omega(\theta)$
and in Lorentz factor $\Gamma_{\infty}(\theta)$, which are extracted from
the simulation, at a time of $t=2.25\,\mathrm{s}$ after launching of the
jet.  To compute the energy density $e$ in the inertial frame for a
perfect fluid moving with Lorentz factor $\Gamma$ we can make use of the
fact that it possesses a conserved quantity, namely, the Bernoulli
constant given by~\citep[see, \eg][Secs. 3.6.2 and
  11.9.2]{Rezzolla_book:2013}
\begin{equation}
\label{eq:bernoulli}
h u_t = \left(\frac{e+p}{\rho}\right) \Gamma = \left(1 +
4\frac{p}{\rho}\right) \Gamma = {\rm const.}\,.
\end{equation}
Hence, for a perfect fluid obeying an ideal-fluid equation of state with
adiabatic index $\gamma=4/3$, expression~\eqref{eq:bernoulli} can be
inverted to obtain the energy density of the fluid in the inertial frame
\begin{equation}
  \label{eq:ee}
  e(r,\theta) = \rho \bigg(1+4\ \dfrac{p}{\rho}\bigg) \Gamma^2 -p\,.
\end{equation}
To obtain the distribution of jet energy per solid angle at latitude
$\theta$, we integrate Eq.~\eqref{eq:ee} over the radial direction:
\begin{equation}
\frac{{\rm d} E_{\rm jet}}{{\rm d}\Omega}(\theta) :=
\int^{r_\mathrm{out}}_{r_{\mathrm{ej,front}}} e(r,\theta)\, r^2 dr\,,
\label{eq:Ee}
\end{equation}
which we will refer to as the ``$E$-structure''\footnote{Indeed,
according to Eq.~\eqref{eq:Ee}, the total energy contained in a 3D volume
$r \in [r_{\mathrm{ej,front}}, r_\mathrm{out}]$ and $(\theta, \phi) \in
\Omega_1$, where $\Omega_1$ is a solid angle, will be: $E_{\rm tot} =
\int_{\theta, \phi \in \Omega_1} ({dE}/{d\Omega}) d \Omega =
\int_{\theta, \phi \in \Omega_1}
\int^{r_\mathrm{out}}_{r_{\mathrm{ej,front}}} e(r,\theta)\, r^2 dr \sin
\theta d \theta d \phi$, as expected from the definition of $e(r,
\theta)$, where we used $d\Omega := \sin \theta d \theta d \phi$.}, where
$r_{\mathrm{ej,front}}$ is the ejecta front, a choice made to avoid
including plasma from unshocked ejecta in the jet structure and is
calculated by :
\begin{equation}
r_{\mathrm{ej,front}} > R_{\rm out} + v_{\mathrm{ej, max}}\, t_{\rm ex} \,,
\end{equation}
where we set $v_{\mathrm{ej, max}} = 0.6$ [Eq.~\eqref{eq:uu}] and $t_{\rm
  ex} = 2.25\,{\rm s}$ is the extraction time of the jet structure in the
simulation domain (see also Sec.~\ref{sec:structure}). The choice of the
definition of the energy structure as the energy per unit volume is
guided by the fact that this function is constant in a top-hat jet, or
generally in an isotropic expansion, and because it is usually considered
in prior numerical work and analytical work.
In this integral, the choice of bounds in radius $r_1$ and $r_2$ is done to
consider only relativistic material in the domain, as explained below.

We define the jet structure in Lorentz factor via the asymptotic Lorentz
factor $\Gamma_\infty$, which can be easily calculated using again the
Bernoulli constant to obtain:
\begin{equation}
h_{\rm{\infty}} \Gamma_{\rm{\infty}} = \Gamma_{\rm{\infty}} = \Gamma
\bigg(1+4\ \dfrac{p}{\rho }\bigg)\,,
\label{eq:lf}
\end{equation}
that is, the Lorentz factor that the material would reach if all its
internal energy were converted to kinetic energy. Then, we determine the
average asymptotic Lorentz factor in the solid angle around latitude
$\theta$
\begin{equation}
\Gamma_{\infty} (\theta) := \frac{\int^{r_\mathrm{out}}_{r_{\mathrm{ej,front}}} \Gamma_\infty(r,
  \theta) r^2 dr}{\int^{r_\mathrm{out}}_{r_{\mathrm{ej,front}}} r^2 d r}\,,
\label{eq:lfavg}
\end{equation}
that we refer to as the ``$\Gamma$-structure''.

We note that when performing the volume integrals to determine both the
$E$- and $\Gamma$-structures we need to avoid taking into account plasma
that is not relativistic a simply part of the moving ejecta. To this
scope, we define different criteria to examine whether a cell should be
excluded or not from the domain. In essence, they consist of cut-offs in
asymptotic velocity, which are considered to only include in the
relativistic jet material with Lorentz-factor high enough to accelerate a
population of non-thermal particles in the forward shock susceptible to
produce synchrotron radiation \citep{2013ApJ...771...54S}.  (see
Sec.~\ref{sec:jet} for more details on these Lorentz-factor cutoffs and
the robustness of our results with respect to different choices).

The procedure outlined above allows us to determine the structure of the
jet, i.e., the dependence of the Lorentz factor and energy of the outflow
on the polar angle with respect to the jet axis. Once emerged from the
jet, the outflow will then decelerate by colliding with the interstellar
medium leading to the afterglow emission. Given the likely achromaticity
of afterglows of jets viewed off axis, we restrict our study in the
emission in the radio band \citep{2019A&A...631A..39D,
  2020MNRAS.493.3521B} and model the afterglow light-curves in the
radio by following the procedure described by \citet{2002ApJ...570L..61G}
and \citet{gill2018}, which we briefly review below.

First, we assume that each angular slice of the relativistic jet evolves
independently of the others; this amounts to ignoring the lateral
spreading (\ie considering $v^\theta=0=v^\phi$) and tends to an
over-estimation of ${\rm d}E_{\rm jet}/{\rm d}\Omega$ and $\Gamma_{\infty}$.
The interaction of the jet with the ambient medium obviously
leads to what is normally referred to as the ``forward'' shock and which
is responsible for dominant part of the emission (see
\citealt{2002ApJ...568..820G} for the validity of this assumption). The
dynamics of the forward shock proceeds through two different phases in
time. First, during the so-called ``coasting phase'', the shock front
propagates with a constant Lorentz factor $\Gamma(r) = \Gamma_0$ until a
deceleration radius
\begin{equation}
  R_d \approx 1.3 \times 10^{17} \bar{n}^{-1/3}
  \left(\frac{E}{10^{53}\,{\rm erg}}\right)^{1/3}
  \left(\frac{\Gamma_0}{100}\right)^{-2/3} \mathrm{cm}
  \,,
\end{equation}
for uniform circum-burst medium, where $\bar{n}$ is the number density of
the ambient medium the jet penetrates after the breakout. We determine
the deceleration radius and subsequent deceleration dynamics for all the
angular slices according to the $E$- and $\Gamma$-structure as described
previously.

Following the case of GW170817 and in line with the low densities
expected in media around BNS mergers, we adopted $\bar{n} = 3 \times
10^{-3}\,{\rm cm}^{-3}$ with a uniform circum-burst medium for all our
afterglow calculations. Once sufficient material has been swept up and
energised in the forward shock, the jet decelerates significantly, such
that, for any polar angle $\theta$, the Lorentz factor evolution with
radius follows ~\citep{Panaitescu_2000}:
\begin{equation}
  \Gamma(\xi) = \dfrac{\Gamma_0 +1}{2}\xi^{s-3}\bigg[
    \sqrt{1+\dfrac{4\Gamma_0}{\Gamma_0+1}\xi^{3-s} +
      \left(\dfrac{2\xi^{3-s}}{\Gamma_0+1}\right)^2} -1 \bigg]\,,
\end{equation}
where $\xi := {r}/{R_d}$. After the coasting phase, the following
relation is recovered for the radial dependence of the Lorentz factor of
the forward shock: $\Gamma (\xi) \propto
\xi^{-(3-s)/2}$~\citep{Blandford-McKee1976}, where $s$ is the power-law
index of the circum-burst medium, equal to $s = 0$ in our case of a
uniform medium. This corresponds to the ``deceleration phase''. To
capture the entire evolution of the afterglow light-curve, we consider
$\xi$ in the range from $r = 0$ up to the Newtonian radius $R_{_{\rm N}}
:= \Gamma_0^{2/(3-s)} R_d$, where the jet becomes non-relativistic.

\begin{figure*}
\centering
\includegraphics[width=0.9\textwidth]{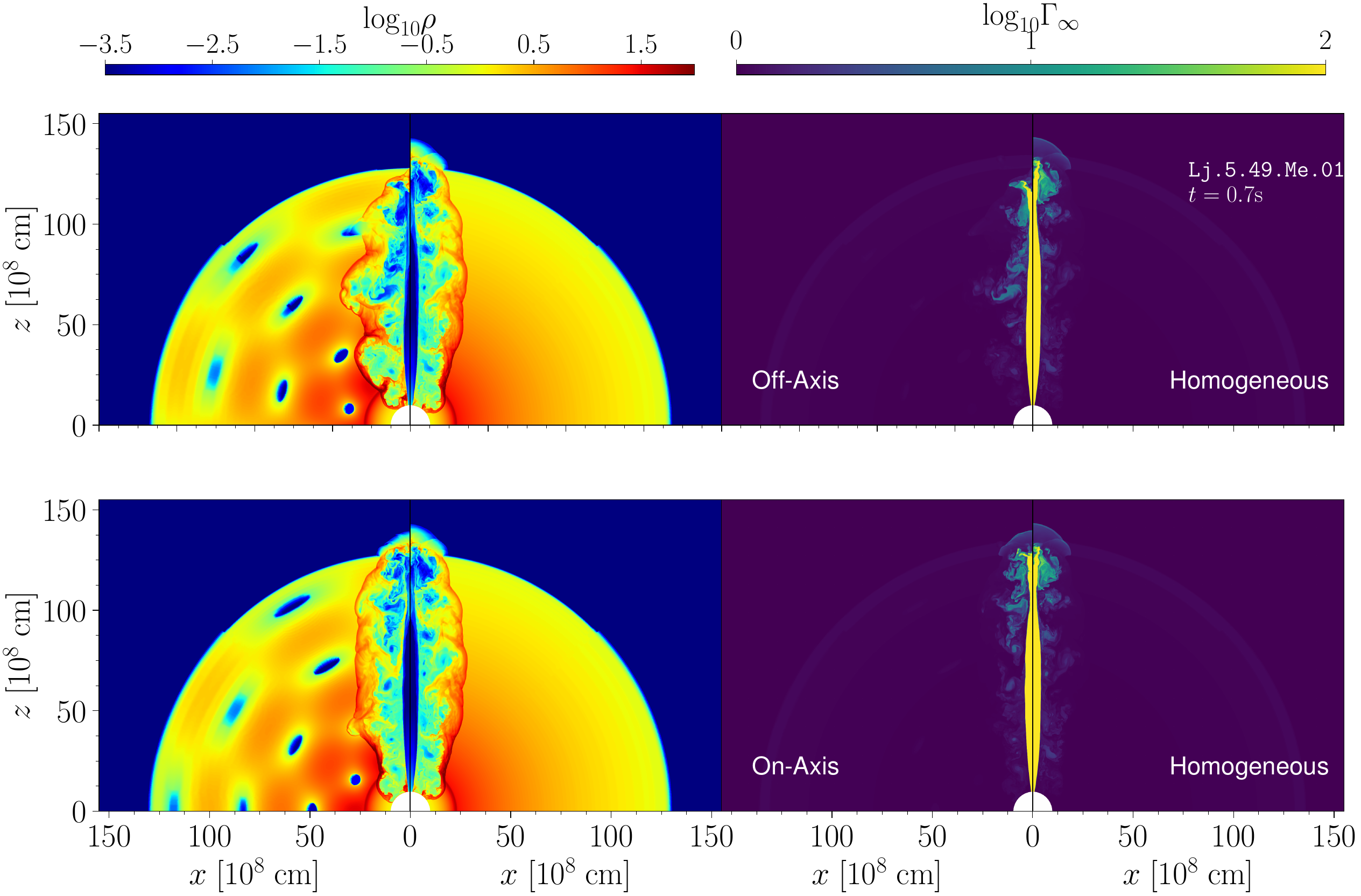}
\caption{Comparison of jet-breakout dynamics for the case
  \texttt{Lj.5.49.Me.010} (\ie $L_{\mathrm{jet}}=5\times 10^{49}\,{\rm
    erg/s}$ and $M_{\mathrm{ej}} = 0.10\,M_{\odot}$). The panels in the
  left column show the rest-mass density, while those in the right column
  report the distributions of the Lorentz factor. To highlight the
  contrast among different distributions, the left part of each panel
  reports the anisotropic distribution of rest-mass density (``off-axis''
  in the top row and ``on-axis'' in the bottom row), while the right part
  the dynamics across a homogeneous distribution. All panels refer to a
  time $t=0.7\,{\rm s}$ since the launch of the jet.}
\label{fig:bo}
\end{figure*}

For a given observation angle $\theta_\mathrm{obs}$ and a given angular
slice of the jet $\theta$, we calculate the afterglow flux assuming
synchrotron emission from the forward shock formed as the relativistic
jet decelerates \citep{2002ApJ...579..699N}. According to the
semi-analytical solution for the deceleration of the jet where the
forward shock forms, the forward shock has a comoving internal energy
density of ${u^\prime} = (\Gamma -1) {n^\prime}m_p$, where ${n^\prime}(r)
\approx 4 \Gamma(r) \bar{n}$ is the comoving particle number density in
the shock and $\Gamma(r)$ is the Lorentz factor of the forward shock once
it has reached radius $r$ \citep{Blandford-McKee1976}.

The fraction of this energy that is deposited to relativistic electrons
is denoted as $\epsilon_e$ and essentially determines the mean Lorentz
factor of these electrons in the frame comoving with the shock. We assume
that the electrons obey a power-law distribution ${n^\prime}(\gamma_e)
\propto \gamma_e^{-q}$, as it is expected from acceleration processes,
and shown by particle-in-cell simulations~\citep[see,
  \eg][]{2011ApJ...726...75S, Meringolo2023}. The power law index
$q\ (>2)$ of the non-thermal electrons in the afterglow emission is
chosen to be $q = 2.2$, as suggested by the ratio of afterglow fluxes in
various bands in GW170817 \citep[e.g.][]{abbott2017b} and other SGRB
afterglows \citep{Cenko_2011}. Furthermore, in the shocked material, a
small-scale turbulent magnetic field is produced and we assume that a
fraction $\epsilon_{_B}$ of the comoving internal energy is transformed
into magnetic energy in the shocked matter to power the synchrotron
emission, such that the magnetic-field strength is $B^2 = 8\pi
\epsilon_{_B} u^\prime$~\citep{Granot_1999}. Hereafter, and unless stated
otherwise, we set $\epsilon_e=0.112$ and $\epsilon_{_B} = 10^{-3}$; these
values are in agreement with the posterior bounds given
by~\citet{Ghirlanda2019}.

Once the distribution of non-thermal electrons and the magnetic-field
strength is set, we can determine the comoving spectral luminosity
$L'_{\nu'}$ due to the synchrotron emission for each angular slice of the
jet over the region of deceleration. We use the expression
for $L'_{\nu'}$ from \citet{gill2018}, equations 13 and following. This
expression covers all the segments of the standard synchrotron spectrum
as a function of the basic physical quantities of the shock that we
previously presented, namely, $n'$, $u'$, $\epsilon_e$, $\epsilon_{_B}$,
and $q$.
  
Finally, we calculate the total flux from the jet at radio frequency $\nu
= 3\,{\rm GHz}$ for a distant observer at viewing angle $\theta_{\rm
  obs}$ and observing time $t$ by integrating the comoving spectral
luminosity over the jet structure on equal-arrival-time surfaces, as
in~\citet{gill2018}. In the end, the jet-structure integral is
written as an integration over angles
\begin{equation}
F_\nu(t) = \dfrac{1}{16 \pi^2 d_{_L}^2}\int \delta_{_D}^3 \,\,
{L^\prime}_{{\nu}^\prime} \,\, d\cos\theta \,d\phi \,,
\end{equation}
where $\delta_{_D} := [\Gamma (1-v_{\rm jet} \cos \chi)]^{-1}$ is the
so-called Doppler factor ($v_{\rm jet}$ is the jet velocity), so that the
two frequencies are related through $\nu = \delta_{_D} \nu'$, and $\chi$
is the angle between the line of sight and the angular slice at $(\theta,
\phi)$ at hand in the integration over the structure. For definiteness,
we chose a luminosity distance of $d_{_L} = 40\,{\rm Mpc}$, small enough
to allow us to ignore redshift corrections. The choice of this distance
is not very important, as we are mostly interested in the morphology of
the afterglow light-curves. Given the small external density we consider
($\bar{n} = 3 \times 10^{-3}\,{\rm cm}^{-3}$), the effects of synchrotron
self-absorption can be neglected \citep{2020A&A...639A..15D}.

\section{Results} 
\label{sec:results}

\subsection{Jet propagation and breakout}
\label{sec:jet}

\begin{figure*}
\centering
\includegraphics[width=0.9\textwidth]{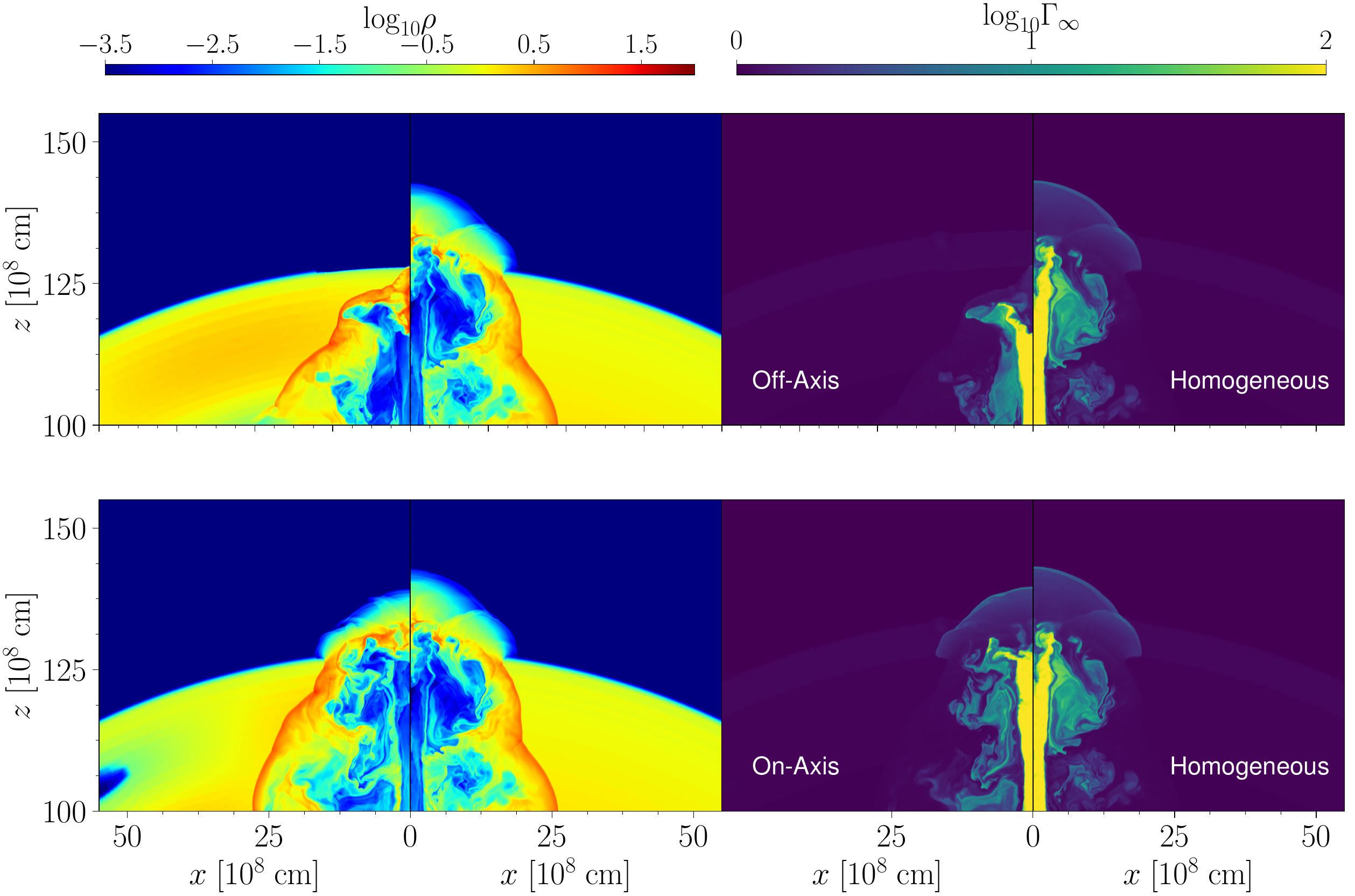}
\caption{The same as in Fig.~\ref{fig:bo} but on a scale highlighting the
  breakout region.}
\label{fig:bo_zoom}
\end{figure*}

We start by studying the effect of the anisotropies in the rest-mass
density distribution on the propagation of the jet in the
ejecta. Figure~\ref{fig:bo} shows a snapshot of the jet-ejecta evolution
at the time of the jet breakout for different ejecta density profiles
from the \texttt{Lj.5.49.Me.010} runs (see Table \ref{tab:runs}). In
particular, Fig.~\ref{fig:bo} reports in its left panel the rest-mass
density distribution, while on the right panel the distribution of the
Lorentz factor (Figure~\ref{fig:bo_zoom} reports the same quantities as
Fig.~\ref{fig:bo} but on a scale that highlights the breakout
region). For each panel, in addition, the left part refers to a
simulation with an anisotropic and inhomogeneous distribution
(``off-axis'' on the top row and ``on-axis'' on the bottom row), while
the right part provides the evolution in the reference case of an
isotropic but inhomogeneous distribution.

When comparing the different parts of the figure, it becomes apparent
that the jet propagation is only mildly affected by the presence of the
anisotropic perturbations in the ejecta distribution. In turn, the
breakout time and energies of the jet at breakout are rather similar
among all cases considered, including the homogeneous one (see discussion
below for details). At the same time, differences do appear in the
morphology of the matter that interacts with the jet and that is set in
turbulent motion by the propagation of the jet. This is more evident in
the case of the ``off-axis'' perturbations, which lead to a significant
lateral expansion of the jet as it encounters regions in the ejecta with
significantly smaller densities. However, differences are present also
for the ``on-axis'' perturbations but, interestingly, these mostly cancel
out. In particular, the acceleration following the interaction with an
underdense region along the jet direction are compensated by the
deceleration when the jet encounters an overdense region. As a result,
the jet breaks-out almost at the same time as in the case of a
propagation across an isotropic ejecta distribution (compare the left
and right parts of the bottom-left panel of Figs.~\ref{fig:bo} and
\ref{fig:bo_zoom}).

\begin{figure*}
\centering
\includegraphics[width=.32\textwidth]{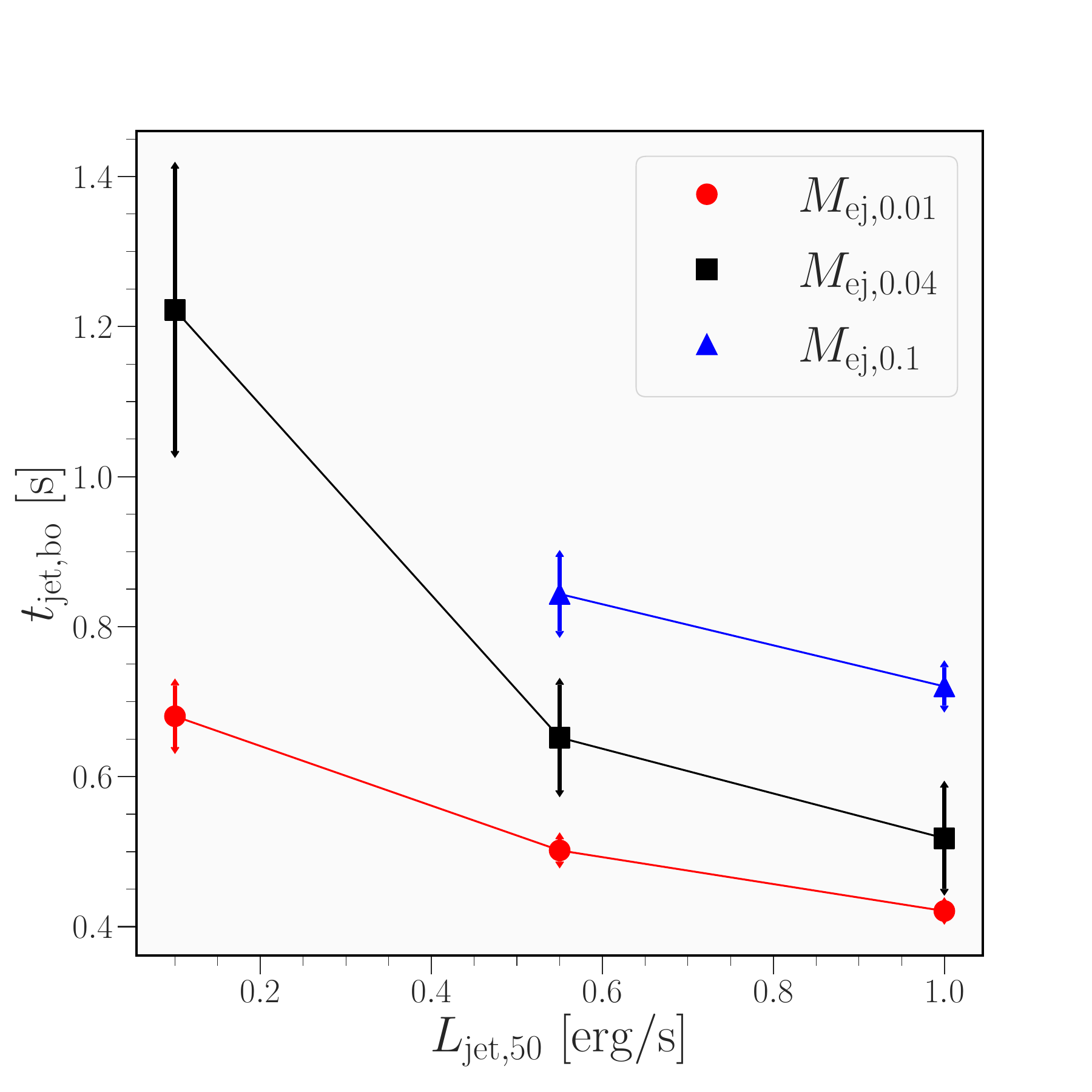}
\includegraphics[width=.32\textwidth]{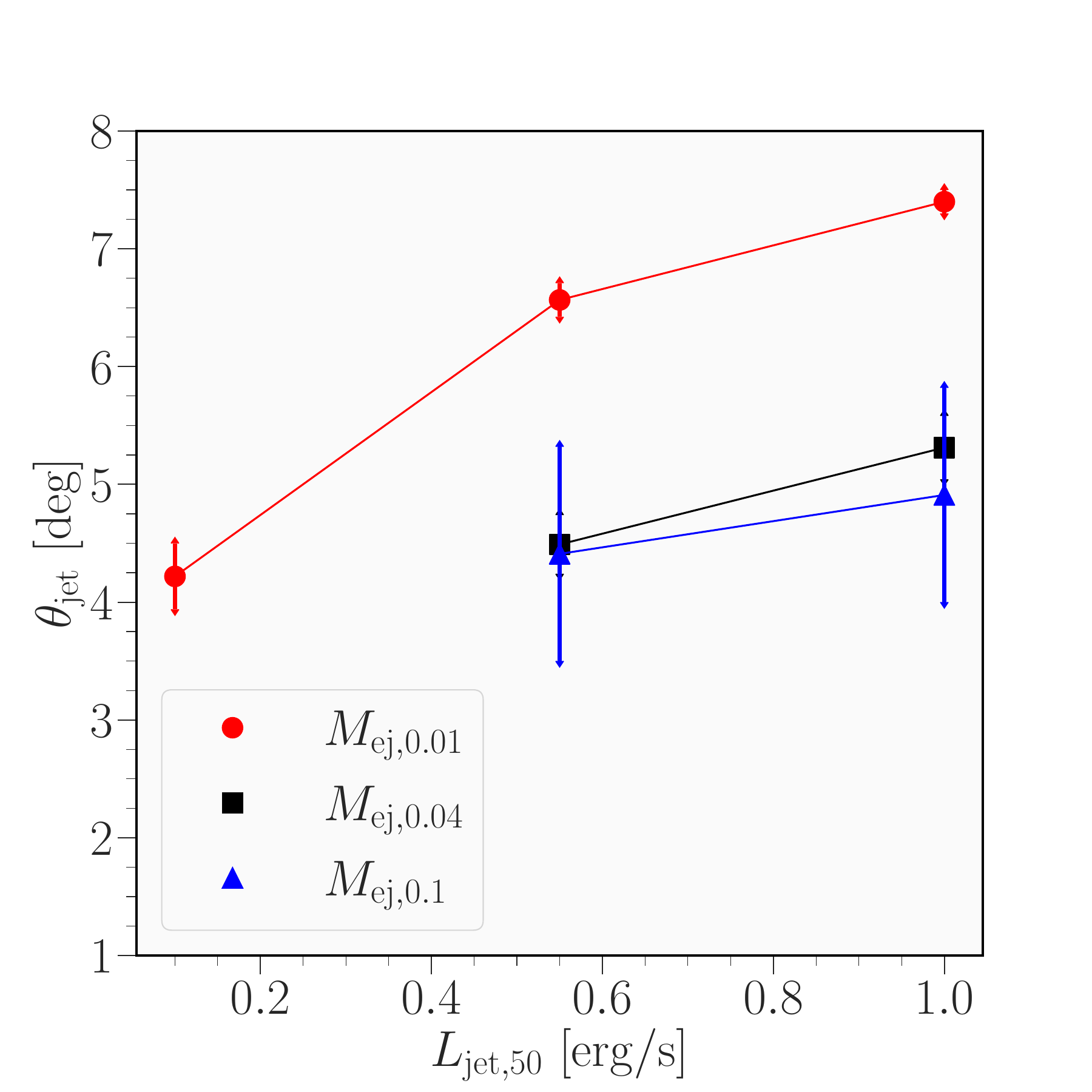}
\includegraphics[width=.32\textwidth]{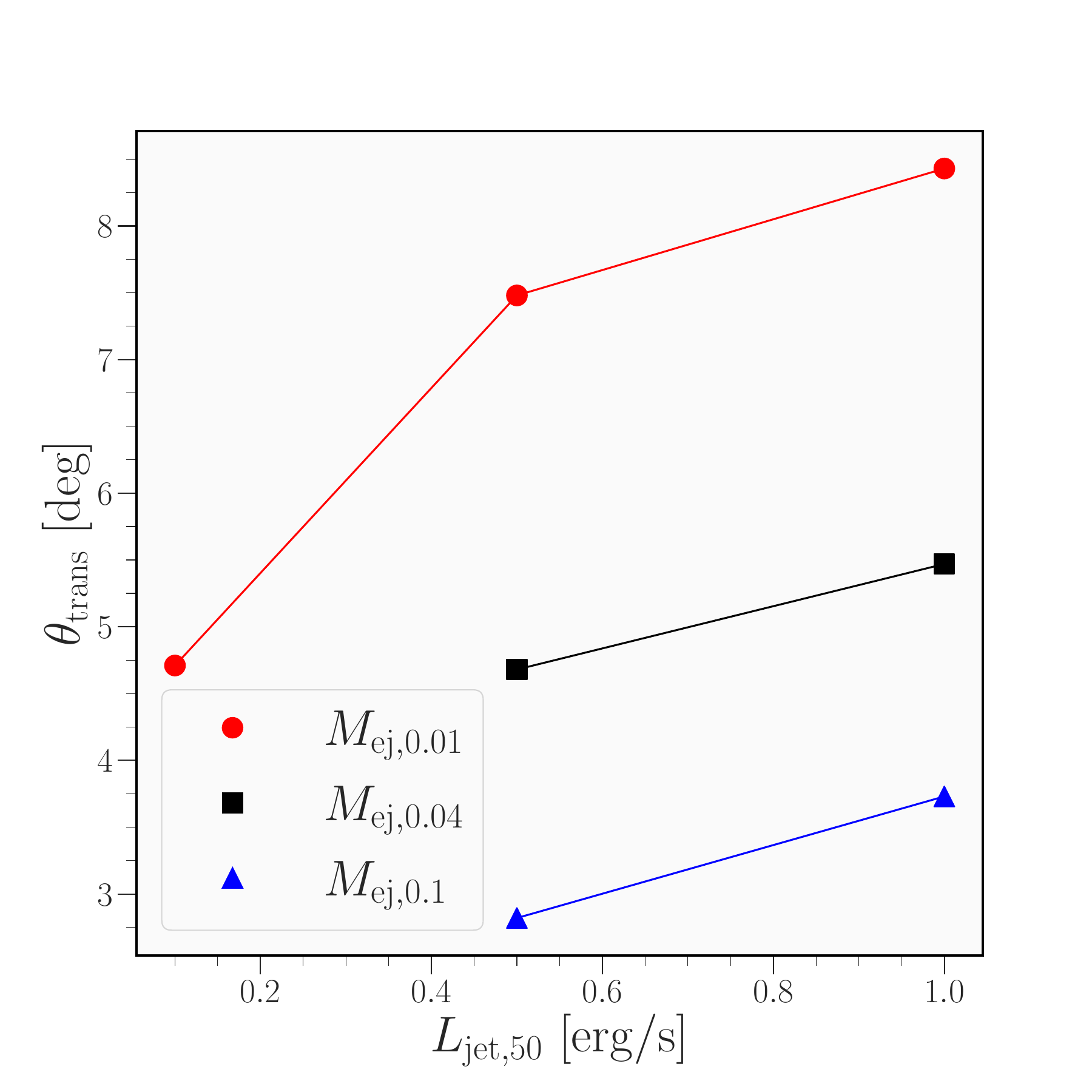}
\caption{\textit{Left panel:} Jet-breakout time from the ejecta as a
  function of the jet luminosity (horizontal axis) and of the mass in the
  ejecta (different coloured lines and symbols). \textit{Middle panel:}
  The same as in the left panel but for the jet opening
  angle. \textit{Right panel:} The same as in the left panel but for the
  transition angle from the steep to the shallow segment in the energy
  polar structure of the jet [see Eq.~\eqref{eq:thetae} for details]. In
  all panels, the vertical bars mark the variation that comes from the
  different rest-mass density distributions.}
\label{fig:bo_full}
\end{figure*}

The qualitative description offered by Figs.~\ref{fig:bo} and
\ref{fig:bo_zoom} can be complemented by a more quantitative description
summarised in Fig.~\ref{fig:bo_full}, which reports, in the left panel,
the jet breakout time $t_{\rm jet,bo}$, and in the middle panel, the
opening angle of the structured jet $\theta_{\rm jet}$. More
specifically, we define $t_{\rm jet, bo}$ as the time when the material
shocked by the propagating shock first overtakes the radially expanding
ejected matter.

\begin{figure*}
\centering
\includegraphics[width=0.9\textwidth]{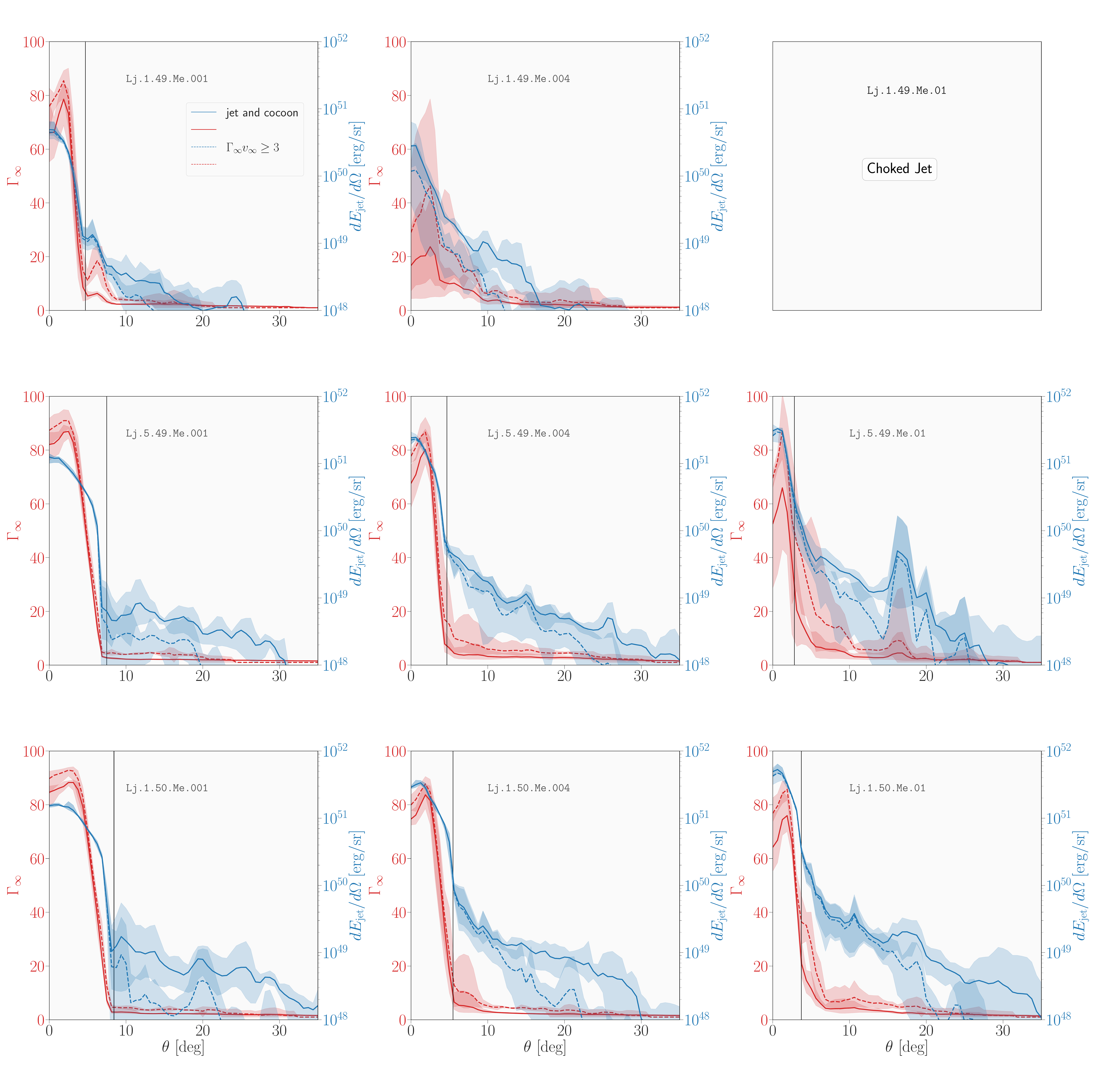}
\caption{Large-scale polar structures at $t=2.25\,\mathrm{s}$ of the nine
  jets simulated and shown in terms of the total energy (blue lines) and
  of the Lorentz factor (red lines). Each panel is marked by the name of
  the corresponding case reported in Table~\ref{tab:runs} and the
  top-right panel refers to the case when the jet is unable to break out
  from the ejecta (\ie ``chocked jet''). Solid lines refer to polar
  structures computed considering the entire plasma outside the ejecta
  (\ie ``jet and cocoon''), while the dashed lines refer to structures
  after imposing a a velocity cut (\ie $u_{\infty} > 3$). In all cases,
  the blue/red shaded areas report the variance due to the different
  types of anisotropy in the ejecta. Shown with a black solid vertical
  line is the transition angle $\theta_{\rm trans}$ (\cf
  Fig.~\ref{fig:bo_full}).}
\label{fig:prof}
\end{figure*}

As we will show in more detail in Fig.~\ref{fig:prof}, the Lorentz factor
experiences a steep decrease outside of inner region (core) that allows
us to robustly define the angular size of the ultra-relativistic core of
the jet, denoted by $\theta_{\mathrm{jet}}$ and defined as the smallest
angle for which $\Gamma_\infty < 10$.  We extract this data from the profiles
presented in Fig.~\ref{fig:prof}. We plot this angle on the middle panel
of Fig.~\ref{fig:bo_full}. We find that this angle varies by at most
$1.5\,\mathrm{deg}$ when changing the pattern of the perturbations in the
ejecta density and, in cases where the ejecta mass is low, this variation
is much less than $10\%$. This suggests that, under the condition of a
successful jet breakout, the jet core is largely unaffected by the ejecta
profile.

The jet break-out and core opening angles reported in
Fig.~\ref{fig:bo_full} are shown as function of the three luminosities
assumed for the jet (horizontal axis) and of the rest-mass density
contained in the ejected material (black, red and blue symbols,
respectively). Furthermore, the variance shown with vertical lines across
the solid symbols reflects the different geometry of the anisotropies
(\ie ``off-axis'', ``on-axis'', or ``mixed'') and indicates one of the
most important results of our work, namely that the impact of the
geometric anisotropies on jet break-out and jet core properties is not
very large and that these jet properties are mainly set by the energy in
the jet and by the mass in the ejecta.

Starting from the left in Fig.~\ref{fig:bo_full}, it is apparent that the
breakout time decreases as the energy of the jet is increased, but also
that it increases as the ejected matter is more massive. The
interpretation of this behaviour is straightforward: the jet travel-time
across the ejecta will depend on its energy and on the matter resistance
it will encounter, so that the most massive ejecta may actually prevent
the breakout of the weakest jet, while the most powerful jet will cross
the least massive ejecta in only $0.4\,{\rm s}$. At the same time, for a
given mass of the ejecta, the breakout time can even double between the
least and the most powerful jets. Interestingly, the jet with $L_{\rm
  jet,50}=0.10\,{\rm erg/s}$ (simulation \texttt{Lj.1.49.Me.010}) is
indeed chocked when going across the massive ejecta with $M_{\rm
  ej}=0.010\,M_{\odot}$. This behavior is in agreement with the
predictions for the ``early-breakout'' criterion of~\citep{Duffell2018}
[Eq.~(20) there] but in tension with the ``late-breakout'' criterion,
which instead predicts a successful jet; as pointed out
by~\cite{2022MNRAS.517.1640G}, this tension may be due to the way that
the jet is being ``turned off''. An abrupt shutdown of a jet, or an
extended transition phase will have an impact on the cases of
``late-breakout''. We also note that the variability of the
  break-out time induced by the anisotropies is larger for the
  intermediate mass ejecta (see, \eg Fig.~\ref{fig:bo_full}). This
  behaviour, could be due to a number of potential effects but we
  conjecture it is the result of the fact that in the case of
  intermediate mass ejecta the system is intrinsically closer to a
  near-cancellation of effects, which is not present if the mass of the
  ejecta is very large or very small. Under these conditions, small
  fluctuations, even numerical in nature, can lead to large variations in
  quantities such as the break-out time or the jet opening angle.

Also, the 2D nature of our simulations, inevitably implies the presence
of the so-called ``plug-instability''~\citep{Zhang_2004, Mizuta_2013,
  Lopez-Camara_2013}, which is not really an instability but rather the
artificial piling up of matter ahead of the propagating jet (in 3D this
matter has more opportunities to move in the direction orthogonal to that
of the jet) and hence affects in part the dynamics of the jet
propagation. Fully three-dimensional simulations have shown that the
plug-instability tends to over-estimate the breakout time by introducing
a systematic bias in the breakout time \citep[see Appendix
  of][]{2021MNRAS.503.4363U}, which is likely to affect also our
results. However, we expect that the basic phenomenological trend of the
breakout time as a function of $L_{\rm jet}$ and $M_{\rm ej}$ portrayed
in the left panel of Fig.~\ref{fig:bo_full} to be robust and hence to be
present also in fully 3D simulations.

The physically intuitive behaviour of the breakout time is shared also by
the jet opening angles $\theta_{\rm jet}$, shown in the middle panel of
Fig.~\ref{fig:bo_full}, respectively. In essence, what these panels
highlight is that this angle increases as the energy of the jet is
increased and decreases as the mass in the ejecta is increased. Once
again, this is not surprising as it is natural to expect that more
powerful jets will deposit systematically larger amounts of energy in the
ejecta, leading to larger opening angles at the breakout. Conversely, a
larger ejecta mass will exert more pressure on the propagating jet and
close the core angle before breakout. Less obvious is that this behaviour
should be essentially linear in $L_{\rm jet}$, while it is clearly
nonlinear in terms of $M_{\rm ej}$ (see Fig.~\ref{fig:bo_full}).

\subsection{Cocoon evolution}
\label{sec:coc}

As mentioned in the previous section, the propagation of the jet within
the ejected envelope leads to a turbulent shear layer that is normally
referred to as the ``cocoon'' for its peculiar shape. At the breakout 
time, Figs.~\ref{fig:bo} and \ref{fig:bo_zoom} illustrate the detailed 
morphology of the cocoon, which can be readily identified by the steep 
gradient in density. More precisely, we distinguish the
matter in the cocoon from the jet material by following the conditions
proposed by \citet{2023MNRAS.520.1111H}, that is, we mark as filled with
cocoon-matter cells for which the following conditions are met:
$\Gamma_{\infty} \le 10$, $\Gamma <5$, and $\lvert v_\theta \rvert > 0$,
where the last criterion is imposed to avoid the mis-identification of
ejected material or interstellar medium (for which $\lvert v_\theta
\rvert = 0$) with the matter in the cocoon. Clearly, Fig.~\ref{fig:bo}
shows that the cocoon varies at different vertical heights and hence it
changes during the evolution. More importantly, it is easy to realise
that the different geometric properties of the ejected material do have
an impact on the morphology of the cocoon.

\begin{figure*}
\centering
\includegraphics[width=.34\textwidth]{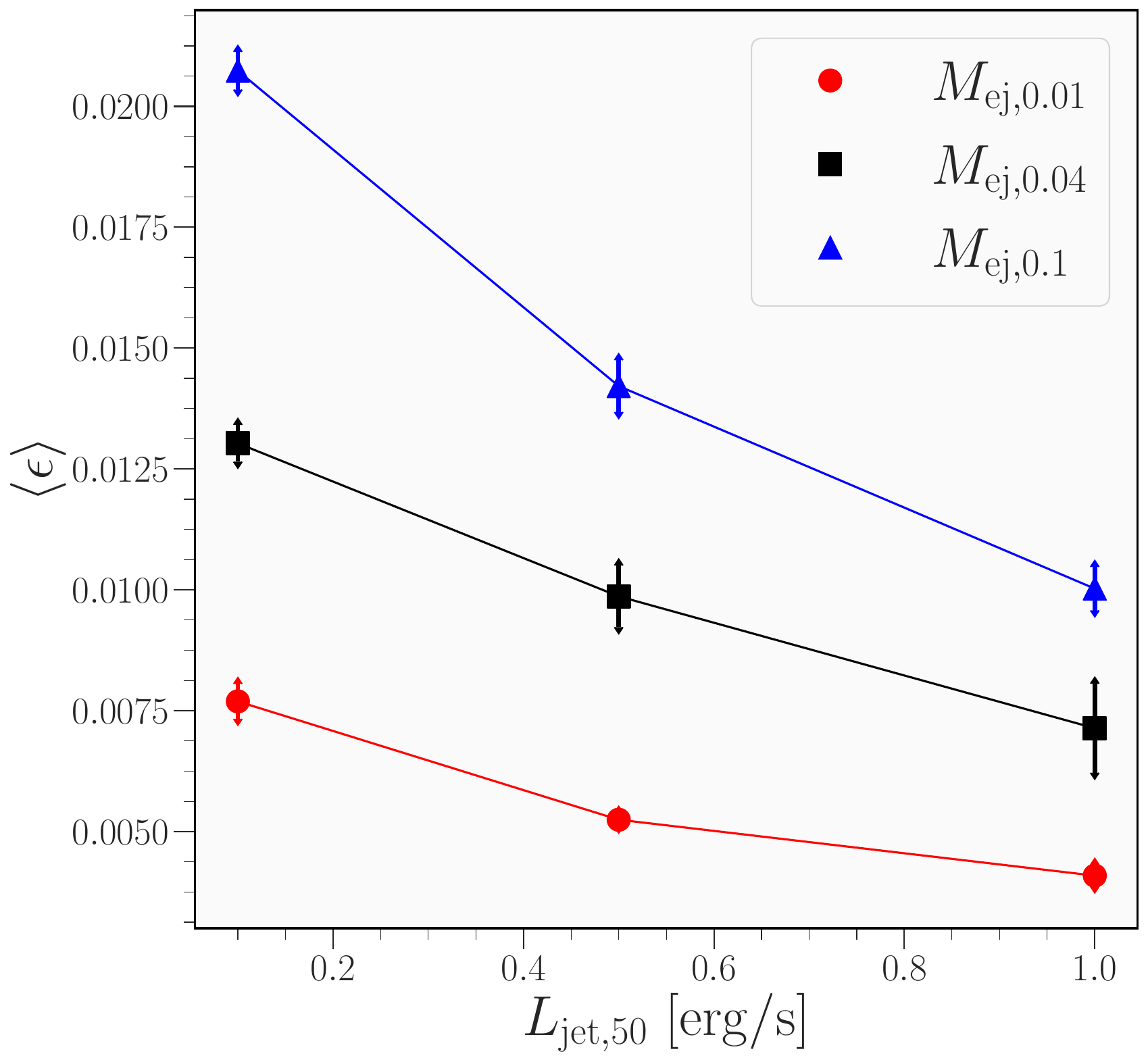}
\includegraphics[width=.32\textwidth]{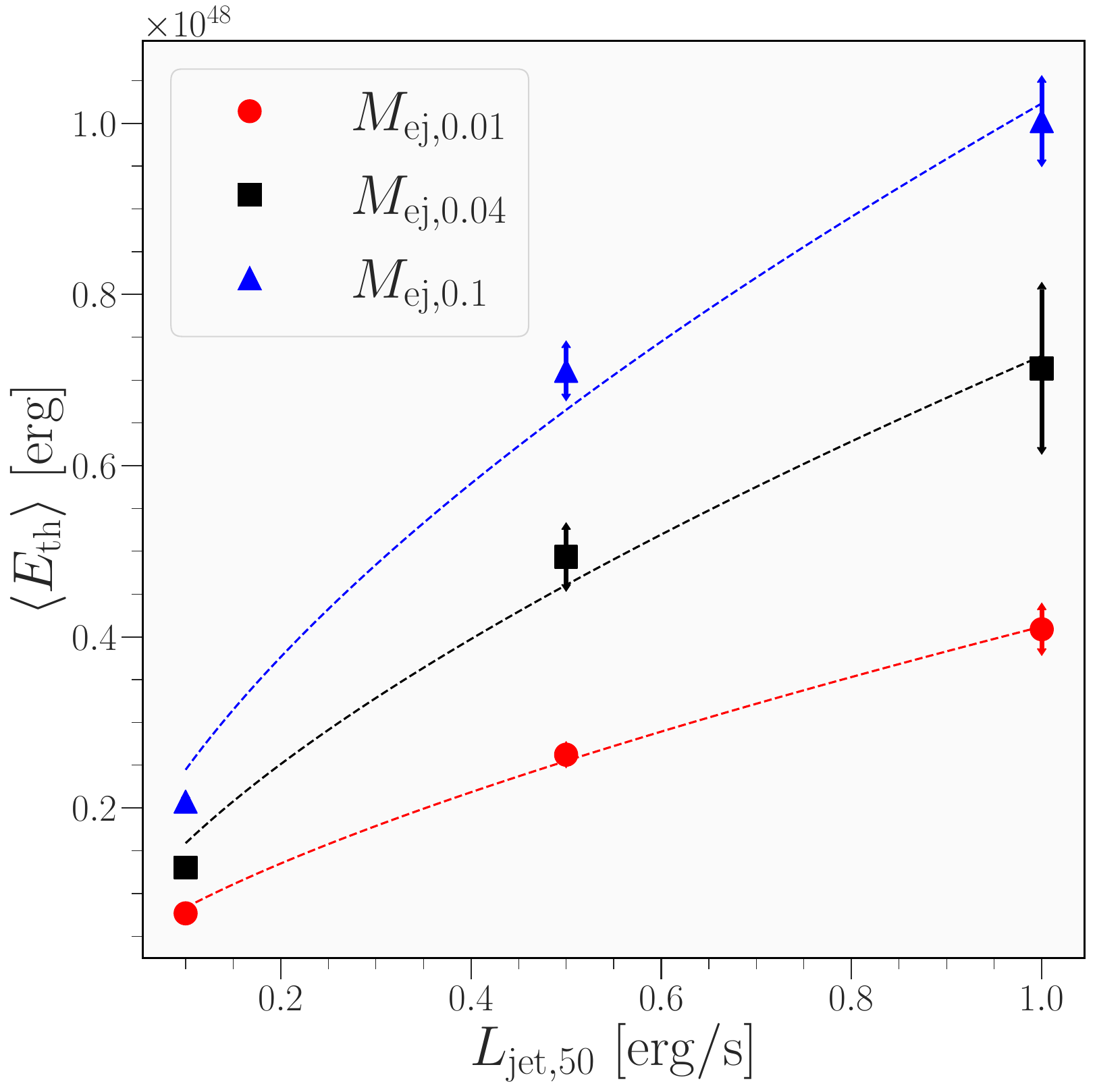}\hfill
\includegraphics[width=.32\textwidth]{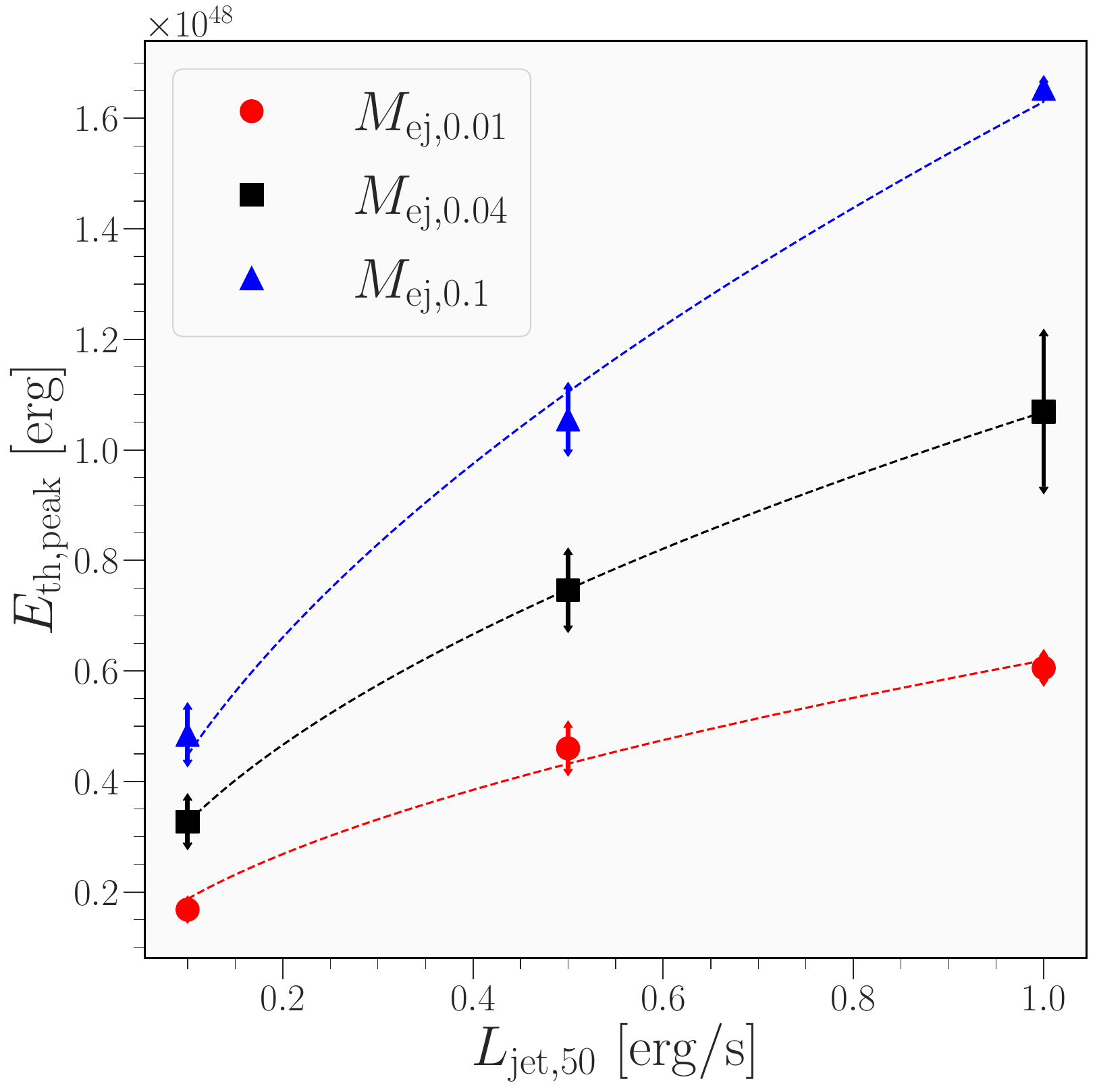}\hfill
\caption{The same as in Fig.~\ref{fig:bo_full} but for the diagnostics of
  the thermal energy in the cocoon during the jet evolution. In
  particular, shown are: the average efficiency of jet-cocoon energy
  transfer [Eq.~\eqref{eq:thermal}, left panel], the time-averaged
  thermal energy of the cocoon up until the jet breakout (middle panel),
  and the peak thermal energy of the cocoon during the jet evolution
  (right panel). In the last two panels, the dashed lines indicate the
  square-root fit with the jet luminosity.}
\label{fig:therm}
\end{figure*}

During its propagation across the ejected material, the jet will deposit
a part of its kinetic energy increasing the internal energy of the ejecta
$E_{\mathrm{th}}$. Since the $L_{\rm jet}$ is the source of energy, we
can define the efficiency $\epsilon$ with which the jet energy is
converted into the thermal energy of the cocoon as
\begin{equation}
\frac{dE_{\mathrm{th}}}{dt} =: \epsilon (t) L_{\mathrm{jet}} \simeq
\langle \epsilon \rangle L_{\mathrm{jet}} \simeq \frac{\langle
  E_{\mathrm{th}} \rangle}{t_{\rm jet, bo}} \,,
\label{eq:thermal}
\end{equation}
where $\langle E_{\mathrm{th}} \rangle$ is the time-averaged thermal
energy deposited in the ejecta from the start and up to the breakout time
$t_{\rm jet, bo}$. The unknown efficiency $\epsilon$ (and its
time-averaged counterpart $\langle \epsilon \rangle$) takes into account
the work done both on the jet head surface via the forward shock and
that done in generating a lateral expansion of the cocoon [the (kinetic)
  energy in the turbulent motion in the cocoon is not included in
  expression~\eqref{eq:thermal}]. Clearly, the value of such an
efficiency will depend on the properties of the ejected material and, in
our setup, on the properties of the anisotropies introduced. Using our
simulations and the measurements of $\langle E_{\mathrm{th}} \rangle$, we
have estimated the average efficiency $\langle \epsilon \rangle$ and
summarised the results in the left panel of Fig.~\ref{fig:therm} for the
configurations considered. It is then interesting to note that the
time-averaged efficiency $\langle \epsilon \rangle$ decreases linearly
with increasing energy in the jet and increases with the mass in the ejecta. Stated
differently, more energetic jets are less efficient in converting the
kinetic energy into thermal energy, especially if the ejecta are not very
massive. This behaviour is rather natural and indeed one would expect a
vanishing efficiency for extremely tenuous ejecta and very energetic
jets, which would propagate in them essentially ballistically. Note also
how the efficiency is also upper-limited by a value of about $1.5\%$,
which corresponds to the configuration \texttt{Lj.1.49.Me.004} that we
will later-on see, is that of a marginally successful
breakout\footnote{Note that in the case of choked-jet configuration
\texttt{Lj.1.49.Me.004} we compute $\langle \epsilon \rangle$ by assuming
$t_{\mathrm{jet,bo}}=t_\mathrm{jet}$.}.

Considering the amount of energy that is needed to explain the early blue
component of the kilonova of GW170817,~\citet{Duffell2018} concluded that
the exchange of thermal energy in the jet-cocoon system is
subdominant. To validate this conclusion, we report in
Fig.~\ref{fig:therm} the time-averaged thermal energy deposited in the
ejecta $\langle E_{\mathrm{th}} \rangle$ (middle panel) and the peak
thermal energy $E_{\rm th, peak}$ (right panel), that is, the largest
value of $E_{\rm th}$ recorded for the cocoon during the jet propagation
through the ejecta, as a function of the energy in the jet and mass in
the ejecta. The behaviour of the two quantities is similar even though
they really refer to distinct quantities as $\langle E_{\mathrm{th}}
\rangle$ contains information on the overall ``heating'' of the ejecta,
while $E_{\rm th, peak}$ is mostly indicative of observational
properties, especially when the observations are performed with
sensitivity-limited instruments. Overall, both panels reveal that the
energy deposited in the ejecta to produce the cocoon increases with the
mass of the ejecta and the luminosity of the jet, and scales like the
square root of the latter, as also anticipated
by~\citep{Duffell2018}. This is shown by the dashed lines that show a
fitting function with power-law index equal to $1/2$.

\begin{figure*}
\centering
\includegraphics[width=.44\textwidth]{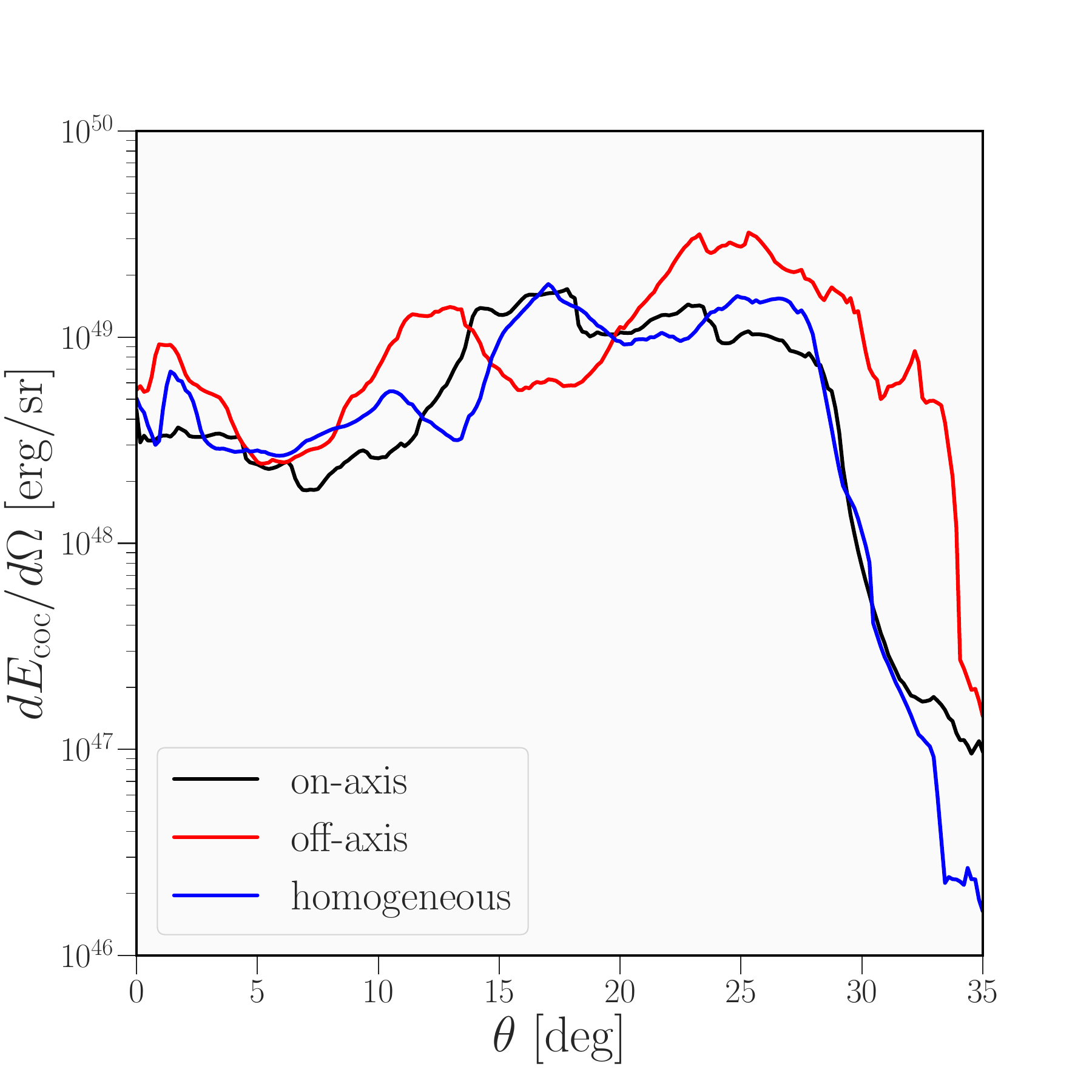}\hfill
\includegraphics[width=.44\textwidth]{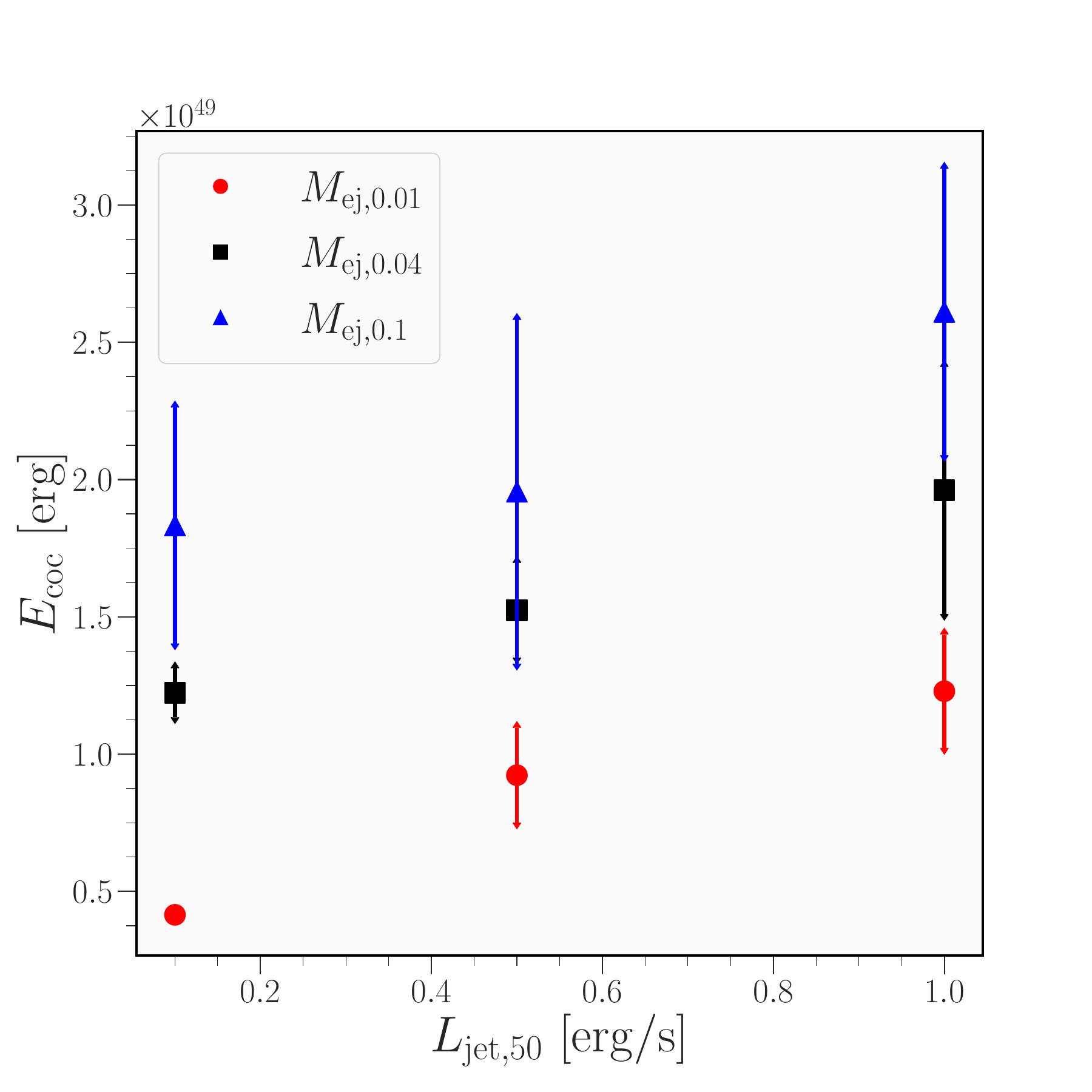}
\caption{\textit{Left panel:} polar distribution of the energy in the
  cocoon for the model \texttt{Lj.1.50.Me.001} and different anisotropies
  (different coloured lines). \textit{Right panel:} total energy in the
  untrapped cocoon shown as a
  function of the jet luminosity and masses in the ejecta (different
  coloured symbols). Also in this case, the vertical bars mark the
  variation that comes from the different density profiles.}
\label{fig:trapped}
\end{figure*}

In addition, note how the variance introduced by the different
anisotropies plays a very modest role in determining the amount of energy
deposited and that the impact is larger for the most energetic
jets. Still within the context of the GW170817 event, early emission from
the cocoon has been studied \citep[see, \eg][]{2017ApJ...834...28N,
  2018MNRAS.479..588G} and associated with early observations of an
electromagnetic counterpart. Because of the delay of about 10 hours in
the localisation of the source, the cocoon contribution to the emission
was likely unobserved. Nonetheless, \citet{2018ApJ...855..103P} claimed
the detection of prompt emission in association with the GW170817
event. More recently, \cite{2023MNRAS.520.1111H} have built an analytical
model to examine the differences between jets produced in collapsars and
in SGRBs environment, the main difference being the static or
homologously expanding ejecta. In this way, they found that at most
$10\%$ of the cocoon mass can escape from the ejecta, and therefore only
a small portion of the cocoon energy can escape and contribute to the
early emission. Indeed, as long as the cocoon is buried in the ejecta and
these are optically thick, it cannot be observed. However, since the
ejecta are expanding and cooling, the optical depth will reach, on a
timescale of about one day, values below unity revealing the
cocoon. Hence, it is important to compare the timescale for the emergence
of the emission from the cocoon with the other timescales, namely $t_{\rm
  jet,bo}$ or the delay between breakout and the $\gamma$-ray emission.

In view of these considerations, and given the wealth of information
provided by our simulations, it is interesting to assess what impact the
anisotropies in the ejecta have on the amount of energy in the cocoon
that is trapped in the ejecta. To this scope, and since all of the ejecta
is already unbound, we need to mark the part of the ``untrapped'' cocoon
as the material of the cocoon that is moving faster than the ejecta;
since the velocity in the ejecta increases linearly with radius
[Eq.~\eqref{eq:uu}] and reaches a maximum of $v_{\mathrm{ej, edge}}$ at
the forward edge, matter in the untrapped cocoon should satisfy the
additional condition
\begin{equation}
\label{eq:matter_coc}
  v_{\mathrm{coc},\infty} \geq v_{\mathrm{ej, edge}}\,,
\end{equation}
where the terminal velocity is calculated as $v_{\mathrm{coc},\infty} =
h_{\rm coc} v_{\mathrm{coc}}$. For this matter, we then measure the
corresponding energy polar distribution ${\rm d}E_{\rm coc}/{\rm
  d}\Omega(\theta)$ of the untrapped cocoon in analogy with what was done
for the energy in the jet [Eq.~\eqref{eq:Ee}]. Also, we compute the total
energy in the untrapped cocoon $E_{\rm coc}$ as the volume integral of
the energy density for matter satisfying the
condition~\eqref{eq:matter_coc}.

The results are summarised in Fig.~\ref{fig:trapped}, whose left panel
reports the angular distribution of the energy of the ``untrapped''
cocoon for three different cases of ejecta, namely, ``on-axis'',
``off-axis'' and ``homogeneous''. Note how the polar distribution in the
cocoon energy is rather similar in the case of the homogeneous and
on-axis ejecta, while it differs in the case of off-axis anisotropies,
where more energy is concentrated at higher latitudes. While this may
appear surprising, it reflects the analogy already remarked between the
propagation across homogeneous and on-axis perturbations, where the
sequence of over- and under-densities along the jet path compensate and
cancel out, hence giving a very similar cocoon and breakout time as in
the homogeneous case (\cf Figs.~\ref{fig:bo} and \ref{fig:bo_zoom}). The
right panel of Fig.~\ref{fig:trapped}, on the other hand, shows $ E_{\rm
  coc} $ as a function of the two main parameters of our setup, \ie
$L_{\rm jet}$ and $M_{\rm ej}$, and the corresponding variance with the
initial anisotropies of the ejecta shown again as vertical bars. Rather
unsurprisingly, the energy in the cocoon that is able to escape grows
with both the energy in the jet and with the mass in the ejecta. In turn,
this implies that even if parts of the ejecta are energised more
significantly because of their lower density, follow-up interactions with
a denser medium do not produce significant changes on the overall
dynamics of the cocoon.

\subsection{Large-scale jet polar structure}
\label{sec:structure}

Using the procedure described in Sec.~\ref{sec:agmeth}, we next discuss
the large-scale polar structure of the jet. For the sake of brevity, we
introduce the short-hand notation $E_\Omega(\theta)$ for the distribution
of jet energy per unit solid angle ${\rm d}E_{\rm jet}/{\rm
  d}\Omega(\theta)$. In particular, for all of the runs performed
Fig.~\ref{fig:prof} reports the polar dependence of the jet Lorentz
factor $\Gamma_{\infty}(\theta)$ (red curves) and energy
$E_{\Omega}(\theta)$ (blue curves). More specifically, for each pair of
$L_\mathrm{jet}, M_{\rm ej}$, the solid lines in the various panels
report respectively the average $E_{\Omega}(\theta)$ and
$\Gamma_{\infty}(\theta)$ across the three perturbation distributions
(``on-axis'', ``off-axis'', and ``mixed''); the shaded areas, on the
other hand, span the ranges $\Delta E_{\Omega}(\theta)$ and $\Delta
\Gamma_\infty(\theta)$ among the three distributions, as due to the
different choices in profiles of the density in the inhomogeneous
ejecta. Note that in each panel of Fig.~\ref{fig:prof} we present the
angular profiles assuming two different cuts in velocity when computing
the energy and Lorentz-factor integrals [Eqs.~\eqref{eq:Ee} and
  \eqref{eq:lfavg}]. The first cut (solid lines), aims at representing
both the cocoon and the jet and corresponds to matter with a velocity
$v_{\infty} > 0.3$. The second cut (dotted line), shows instead only the
mildly-relativistic to relativistic outflow and corresponding to a cut
$\Gamma_{\infty} v_{\infty} \ge 3$. This cut is meant to refer to
material that is expected to contribute the most to the afterglow
radiation through the forward shock. Indeed, below this velocity, the
efficiency of the forward shock in accelerating the population of
non-thermal electrons necessary for the synchrotron emission from the
shocked region, is considerably smaller~\citep[see,
  \eg][]{2012A&A...538A..81M,2019MNRAS.485.5105C}.

Figure~\ref{fig:prof} reveals that, except for the particular case of
\texttt{Lj.1.49.Me.004}, which we discuss separately later, all the
simulations show a similar behaviour for the polar structure of the
jet. In particular, the polar profiles of the Lorentz factor consistently
show a ``core'' with large (\ie $\Gamma_\infty \sim 60-90$) and
near-constant values of $\Gamma_\infty$ within an angle of $\theta
\approx 5\,\mathrm{deg}$, which is closely related to the initial jet
opening angle $\theta_{\rm jet, i}$.  The Lorentz factor then exhibits a
sharp decrease down to $\Gamma_\infty \lesssim 10$ for $\theta \gtrsim
7\, {\rm deg}$; this sharp decrease, which we have indicated with
$\theta_{\rm jet}$ in Fig.~\ref{fig:bo_full}, can be well described by a
power-law dependence $\Gamma_\infty(\theta) \propto \theta^{-\beta}$ with
$\beta > 7$ for all configurations considered.

Similarly, the polar profiles of the energy also consistently exhibit a
core region, with an energy level that is largely independent of the
choice of the perturbation pattern in the ejecta. Outside of this core,
the energy polar structures then present a decreasing segment where, just
like for the Lorentz factor, the profiles are largely independent of the
density anisotropies in the ejecta. However, in contrast to the
$\Gamma$-structure, the $E$-structure does not have a sharp fall-off for
$\theta \gtrsim \theta_{\rm jet, i}$, but a much slower decay ranging
from a flat distribution (e.g., \texttt{Lj.5.49.Me.001}) to slopes
with a power-law index $\beta \sim 3$ (e.g., \texttt{Lj.5.49.Me.004}) up
to large latitudes up to $\theta \approx 25\,\mathrm{deg}$. This region
of slower decay contains a significant reservoir of energy and,
remarkably does depend on the actual pattern of anisotropies in the
ejecta and, as we will show below, strongly impacts the afterglow
emission, most notably at early times.

To discuss this original feature of the $E$-structures, we introduce the
angle $\theta_{\rm trans}$, defined as the transition angle from the
steep segment to the shallow segment in the energy structure of the
jet. To capture this transition from the ejecta-pattern-independent to
the pattern-dependent segment of the structure we define $\theta_{\rm
  trans}$ as the angle marking a relative difference larger than $50\%$,
between the average structure and the structure range \ie
\begin{equation}
\theta_{\rm trans} = {\rm min} \left\{\theta \left| \frac{\Delta
  E_{\Omega}(\theta)}{E_{\Omega}(\theta)} > 50\%\right.\right\}\,,
\label{eq:thetae}
\end{equation}
which depends on the injected jet energy and on the mass of the
ejecta. This angle is reported as a vertical black solid for all cases
shown in Fig.~\ref{fig:prof}\footnote{Note that no $\theta_{\rm trans}$
can be determined for the configurations \texttt{Lj.1.49.Me.004} and
\texttt{Lj.1.49.Me.010}.}. In addition, in the right panel of
Fig.~\ref{fig:bo_full}, we report $\theta_{\rm trans}$ as a function of
$L_{\rm jet}$ for different values of $M_{\rm ej}$. We find that,
analogously to $\theta_{\rm jet}$, $\theta_{\rm trans}$ increases with
$L_{\rm jet}$ and decreases with $M_{\rm ej}$, underlining the larger or
lesser capacity for the jet to expand without being affected by the
anisotropies in the ejecta.

Before concluding this section, we concentrate on the case of
\texttt{Lj.1.49.Me.004}, which, as already anticipated, deviates from
this general trend. While this configuration does produce a large-scale
jet with a significant Lorentz factor in the core (on average in the
core, $\Gamma_\infty \gtrsim 30$), there are some density anisotropies in
the ejecta that do not lead to an ultra-relativistic jet emerging on a
large scale (indeed $\Gamma_\infty \lesssim 10$ in the core for a
``off-axis'' anisotropic distribution). In addition, the average Lorentz
factor in the core is significantly lower than that of the injected jet,
and is very low with respect to constraints on SGRB jets.  Considering
that a similar configuration with a more massive set of ejecta, \ie
\texttt{Lj.1.49.Me.010}, does not lead to a successful breakout but to a
chocked jet, we interpret the phenomenology associated with
\texttt{Lj.1.49.Me.004} as to that of a marginally successful
breakout. In this configuration, therefore, the energy injected is only
marginally sufficient to produce a successful motion across the ejecta
that, when slightly more massive, are instead capable of preventing the
progression of the jet. We can use these two cases, namely
\texttt{Lj.1.49.Me.004} and \texttt{Lj.1.49.Me.010}, to set a minimum
threshold for the ``critical efficiency'' necessary for a successful
breakout given an ejecta mass and a jet propagating at a given
luminosity. In particular, using the data for \texttt{Lj.1.49.Me.010} we
determine such an efficiency to be 
\begin{equation}
  \label{eq:eta_crit}
  \eta_{\rm crit} := 
  \left(\frac{L_{\rm jet}}{10^{50}\,{\rm erg/s}}\right)
  \left(\frac{t_{\rm jet,bo}}{1\,{\rm s}}\right)
  \left(\frac{M_{\rm ej}}{10^{-3}\,M_\odot}\right)^{-1}
  \gtrsim 0.03\,.
\end{equation}
The robustness of this result is provided by the fact that all
configurations have efficiencies larger than $\eta_{\rm crit}$ and the
same behaviour discussed so far is measured when using other extraction
times, \eg $t_{\rm ex} = 2.0,\, 2.2\,{\rm s}$, or when varying the
threshold for our velocity cut, \eg $\Gamma_{\infty} v_{\infty} = 2.0,\,
2.5$. Finally, the estimate in Eq.~\eqref{eq:eta_crit} is in reasonable
agreement with similar but distinct estimates~\citep{Duffell2018}.

\begin{figure*}
\centering
\includegraphics[width=0.9\textwidth]{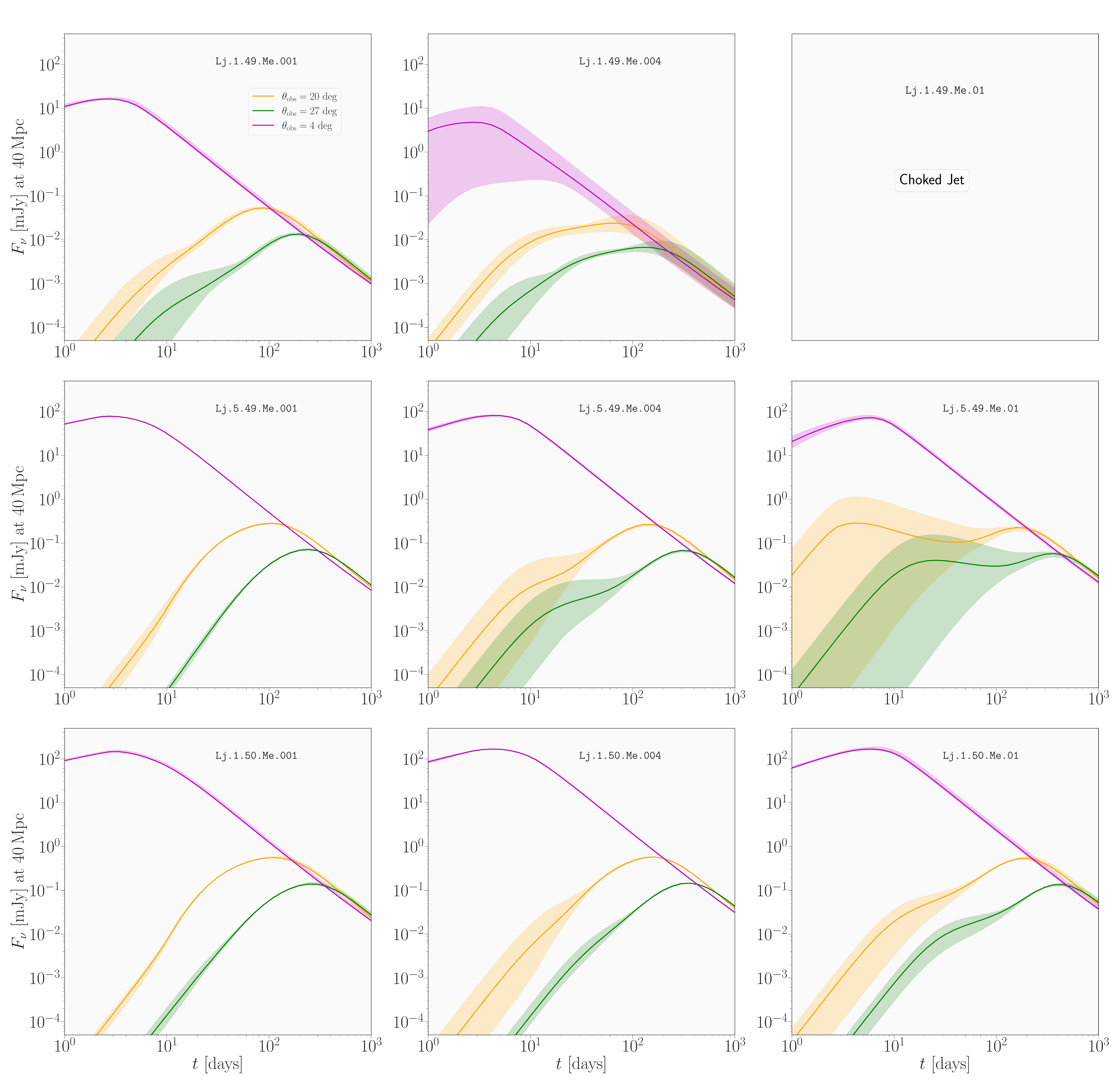}
\caption{Afterglow emission of the nine jets simulated; each panel is
  marked by the name of the corresponding case reported in
  Table~\ref{tab:runs} and the top-right panel refers to the chocked
  jet. For each panel, the different lines refer to the different viewing
  angles of the observers and the corresponding coloured shaded areas
  report the variance due to the different types of anisotropy in the
  ejecta. Note that in the case of model \textit{Lj.1.49.Me.004} the
  light-curve is double peaked as a result of complex distribution of the
  jet energy in the shallow segment (see main text for a
  discussion).}
\label{fig:agsfull} 
\end{figure*}

\subsection{Afterglow predictions and observations}
\label{sec:agres}

Starting from the jet structures shown in Fig.~\ref{fig:prof}, we can
produce the radio afterglow light-curves emission following the procedure
discussed in Sec.~\ref{sec:agmeth}. These light-curves are reported in
Fig.~\ref{fig:agsfull} for three different viewing angles $\theta_{\rm
  obs} = 4$ (magenta lines), $20$ (orange lines) and $27\,{\rm deg}$
(green lines), representing, respectively lines of sight that are either
``on-axis'' (magenta line) or ``off-axis'' (orange and green lines); the
shading around these lines follows the convention already employed in
Fig.~\ref{fig:prof}, namely, marking the variance when considering
different anisotropies in the ejecta. Postponing again the discussion of
the marginally successful breakout \texttt{Lj.1.49.Me.004}, we note that
the afterglows for an observer whose line of sight is on-axis, exhibit a
short brightening phase followed by a monotonous power-law decay with a
by a monotonous power-law decay with a temporal slope of $F_\nu \propto
t^{-1.8}$, which coincides with the theoretical prediction $F_\nu \propto
t^{-3(q-1)/2}$ for the afterglow from a non-expanding jet observed below
the synchrotron cooling break, where $q = 2.2$ is the spectral index of
the population of non-thermal electrons in the forward shock (see
Sec.~\ref{sec:agmeth})~\citep{1998ApJ...497L..17S,
  2018MNRAS.481.2581L}. On these lines of sight, the afterglow is fully
dominated by the emission from the core jet which, as discussed in the
previous section, does not depend on the type of anisotropies in the
ejecta. On the other hand, for observers with off-axis lines of sight,
the afterglows exhibit light-curves as first revealed by GRB170817A, that
is, with a long-lasting brightening phase lasting for up to hundreds of
days, before reaching a broad peak and connecting to the power-law decay
seen by on-axis observers \citep{2018ApJ...868L..11M, gill2018}. This
general behaviour is due to the delayed opening of the beaming cone of
the jet core as a result of the large difference between the line of
sight and the direction of propagation of the jet. Indeed, given the
estimates of the core opening angle $\theta_{\rm jet} \sim 5\,{\rm deg}$
(Fig.~\ref{fig:bo_full}), the off-axis afterglows of
Fig.~\ref{fig:agsfull} have $\theta_{\rm obs} / \theta_{\rm jet} > 4$,
well inside the regime of significantly off-axis lines of sight. On these
off-axis lines of sight, and as time proceeds, the detected radiation is
dominated progressively by decelerating material which is moving with
angles from $\theta \lesssim \theta_{\rm obs}$ down to $\theta \sim 0$,
\ie the jet core.

As discussed above, the jet structures at high latitudes are most
affected by the ejecta density perturbations, explaining the variation
(shown by the shading in Fig.~\ref{fig:agsfull}) in the afterglow
light-curves at early times. As the jet propagates, the variance in the
light-curves reduces as the emission becomes dominated by the core
structure, which is insensitive to the ejecta pattern.

Figure~\ref{fig:agsfull} is also useful to appreciate the impact that the
different anisotropies have on the variance of the light-curves as a
function of the jet luminosity and ejected mass. Recalling that the
smaller the impact, the smaller the variance in the light-curves -- and
hence the width of the shaded regions -- it is apparent that the variance
is reduced as the injected energy in the jet is increased $L_{\rm jet}$
(\ie from top row to bottom row), but it increases with the mass in the
ejecta $M_{\rm ej}$ (\ie from left column to right column). As a result,
the impact of the anisotropies in the ejecta effectively increases with
the ratio $M_{\rm ej}/L_{\rm jet}$. It is also particularly interesting
to note the special behaviour of the afterglow light-curves for the case
\texttt{Lj.5.49.Me.010} (middle row, right column), which exhibit two
local maxima. The origin of these two maxima in the afterglows is
probably to be found in the two local maxima that are present in the
energy distribution for this configuration with a high-mass ejecta (\cf
panel in middle row, right column of Fig.~\ref{fig:prof}) and, in
particular to the significant amount of energy present at high-latitudes,
\ie at $\theta \simeq 15\,{\rm deg} \gg \theta_{\rm trans} \sim 3\,{\rm
  deg}$. The presence of high-Lorentz factor material at these high
latitudes also plays a role in this morphology of the afterglow light
curves.  In particular, it is reasonable that first (earlier) of these
peaks is produced exactly by the shallow portion of the jet energy
structure which dominates at early times and progressively decays. A few
tens or hundreds of days later depending on the observer line of sight,
the emission is instead dominated by the steep jet core and this leads to
the second and more traditional emission observed also in the other
configurations.

Interestingly, \citet{2020MNRAS.493.3521B} have analysed in which
conditions a jet structure with a single power-law in energy and Lorentz
factor can give rise to a double-peaked afterglow, just as observed for
our \texttt{Lj.5.49.Me.010} system. They found that such afterglows are a
generic feature of light-curves from off-axis lines of sight resulting
from the interplay between the progressive debeaming of the core and the
deceleration of the material on the line of sight. While this is indeed a
possibility, our results here show that double-peaks in the light-curves
can also arise from a simple angular exploration of the energy structure
of the jet, provided it has two portions, with a steep inner core and a
shallow exterior with a local maximum at large latitudes. Hence, the
system \texttt{Lj.5.49.Me.010} represents a further illustration of the
degeneracies that can be present when inferring jet structures from
afterglow light-curves; we will come back to this point in
Sec.~\ref{sec:agdis}.

Finally, the presence of an extended shallow segment in the jet structure
for systems with moderate $L_{\rm jet}/M_{\rm ej}$ ratios results in
long-lasting phases of slow brightening of the afterglow, as observed in
the case of GRB170817A, as the emission is dominated by material
progressively ranging up the shallow segment (see. e.g., configurations
\texttt{Lj.5.49.Me.004}, \texttt{Lj.1.49.Me.004}). Indeed, the very
luminous jets emerge with a steep structure (e.g.,
\texttt{Lj.1.50.Me.001}), and display a rapidly brightening afterglow,
close to $F_\nu \propto t^{3}$ which is the case for top-hat
\citep{2002ApJ...579..699N}. On the contrary, systems such as
\texttt{Lj.1.50.Me.010} display a slow phase before the peak, signalling
the shallow segment. Furthermore, in the specific case of
\texttt{Lj.1.49.Me.004}, we have shown in Sec.~\ref{sec:structure} that
the relativistic nature of the jet after breakout was very sensitive to
the perturbation pattern. The structure of the outflow after a marginally
successful or a failed breakout could then be better described as a
nearly spherically symmetric outflow with a radial stratification in
velocity~\citep{2017Sci...358.1559K, 2018MNRAS.479..588G}, rather than as
a jet with angular dependence of energy and Lorentz factor. In this case,
the afterglow radiation is better captured by a model of continuous
energy injection in the mildly relativistic forward shock, as the early
afterglow of GRB170817 suggested~\citep{2018PhRvL.120x1103L,
  2018A&A...613L...1D}, rather than the ultra-relativistic structured jet
afterglow model used here.

\begin{figure}
\centering
\includegraphics[width=1.0\columnwidth]{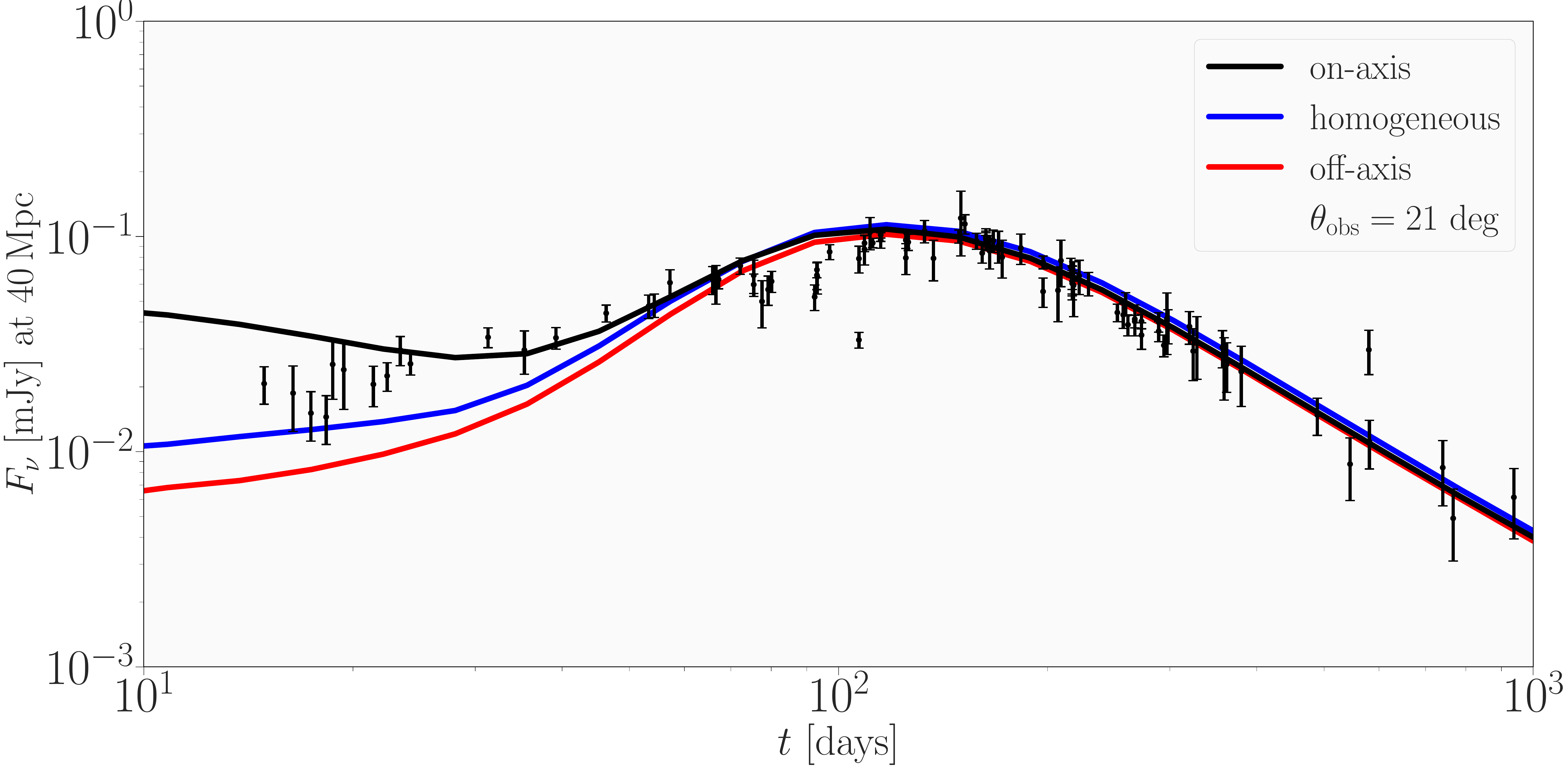}
\caption{Comparison of the radio afterglow from GW170817 (black circles
  with corresponding error bars) with the afterglow light-curves relative
  to model \texttt{Lj.5.49.Me.004} for an observer with viewing angle
  $\theta_{\rm obs} = 21\,{\rm deg}$. The different lines refer to the
  different anisotropies in the ejecta (red line for off-axis, blue line
  for on-axis and black line for the homogeneous distribution). The data
  is taken at a frequency of $\nu = 6\,\mathrm{GHz}$ and up until
  $940\,\mathrm{d}$ after the burst.}
\label{figs:ag_fit}
\end{figure}

A couple of remarks are useful before concluding this section. First,
Fig.~\ref{fig:prof} shows that the central regions of the jet structure
for case \texttt{Lj.1.49.Me.004} can contain material at the edge of our
velocity cut considered for the structure, at $\Gamma_{\infty} v_{\infty}
\sim 3$. This material, in turn, corresponds to the lower bound of the
shaded area for the afterglow light-curve in Fig.~\ref{fig:agsfull}
relative to an on-axis observer. This material, however, is likely not to
be able to accelerate electrons to the level of efficiency of $\epsilon_e
= 0.1$ considered in this work; as a result, the lower bound on the
afterglow light-curve for this system is likely
overestimated. Furthermore, given the high sensitivity of the outcome of
this system on the ejecta anisotropies, it is reasonable to consider the
shading in the afterglow of in Fig.~\ref{fig:agsfull} as extending to
much lower fluxes. Second, for the configuration \texttt{Lj.5.49.Me.004},
we examine how different cases match the radio observations of the
afterglow of GW170817 \citep[data collected from][]{2017ApJ...848L..20M,
  2017ApJ...848L..21A, 2017Sci...358.1579H, 2018Natur.561..355M,
  2018ApJ...858L..15D, 2018ApJ...862L..19N, 2018MNRAS.478L..18T,
  2019ApJ...870L..15L, 2019ApJ...883L...1F, 2019ApJ...886L..17H,
  2021ApJ...922..154M}. More specifically, in Fig.~\ref{figs:ag_fit} we
overlay the data from GW170817 (black circles) with the afterglow
light-curves determined from the jet structures found in the case of
either homogeneous ejecta (blue solid line) or for anisotropic ejecta
that are either on-axis (black solid line) or off-axis (red solid
line). To this scope, we used the jet polar-structure profiles obtained
by applying the velocity cut $\Gamma_{\infty} v_{\infty} > 3$, and
afterglow parameters $\bar{n} = 5\times 10^{-4}\,{\rm cm}^{-3}$,
$\epsilon_e=10^{-0.32}$, and $\epsilon_{_B} = 10^{-3}$, while the viewing
angle is assumed to be that of an off-axis observer with $\theta_{\rm
  obs} = 21\,{\rm deg}$. The values of the afterglow parameters
$\bar{n}$, $\epsilon_e$ and $\epsilon_B$ are different than those adopted
in Fig.~\ref{fig:agsfull}, as they were determined to best-fit the
GW170817 data assuming the structure from the \texttt{Lj.5.49.Me.004}.

The figure clearly illustrates that the afterglows differ mostly at
earlier times (a result we have already anticipated) and that these
differences are rooted both in the anisotropies of the ejecta and in the
emission from high latitudes. In view of these results, and while there
may be some uncertainty in the structure at these high angles, it is
clear that earlier photometry data would be the sharpest way to study
jet-ejecta interaction through its signature in the afterglow (see also
Sec.~\ref{sec:agdis} for a discussion of afterglows to infer jet-ejecta
interaction physics and the particular case of GW170817).

\subsection{On the use of afterglows to study jet structures}
\label{sec:agdis}

The afterglows of relativistic outflows can obviously be used to infer
the structure of these jets. This structure can, in turn, provide insight
on the physics at smaller scales, on the launching of the jet by the
central engine and its interaction with the merger ejecta. Overall, our
simulations show that the interaction of the jet with the ejecta, quite
independently of their geometric anisotropic properties, leads to an
energy polar structure of the jet that is characterised by a steep
segment followed by a shallow segment can arise from jet-ejecta
interaction (\cf Fig.~\ref{fig:prof}), and that the extent of the shallow
segment depends on the strength of this interaction as measured by the
ratio of the ejecta mass and of the jet luminosity, $M_{\rm ej}/L_{\rm
  jet}$ ratio. Interestingly, to our knowledge, this double
steep-shallow-segment jet structure in energy has not been found in
previous study of jet-ejecta interaction, and has not been considered as
a functional form for the fitting of afterglow data. It does, however,
appear as a possible jet structure reconstructed directly from the
afterglow data of GRB170817A~\citep{2021MNRAS.501.5746T}. In particular,
as shown in Fig.~\ref{fig:agsfull}, this structure can give rise to a
variety of afterglow morphologies for off-axis observers, from a
fast-rise-fast-decline for very small interactions resulting in near
top-hat structures (e.g., \texttt{Lj.1.50.Me.001}) to
slow-rise-fast-decline for moderate interaction resulting in a reduced
shallow segment (e.g., \texttt{Lj.1.50.Me.010}) and a double-peaked
afterglow for significant interaction resulting in a large energy
reservoir at high-latitudes (e.g., \texttt{Lj.5.49.Me.010}). Analytical
studies have shown that this variety of light-curve morphologies can also
arise from simple power-law structures \citep{2020ApJ...895L..33B},
underlining once more the large degeneracy (and risks) in reconstructing
structure from afterglow light-curves. As an additional caveat, our
simulations also highlight that differences in the distribution of the
rest-mass density of the ejecta can have a strong impact on the
large-scale structure of the jet and of the cocoon. This makes the
inferring of the ejecta properties more challenging even when assuming a
perfect knowledge of the jet structure. At the same time, the simulations
also show that the impact of the ejecta properties depends systematically
on both the jet luminosity $L_{\rm jet}$ to the ejecta mass $M_{\rm ej}$,
and has a regular behaviour with their ratio.

We should also remark that the effect of the jet energy structure has
also been recently considered on slightly misaligned lines of sight,
i.e., with $\theta_{\rm obs} \lesssim 2 \times \theta_{\rm jet}$, in relation to
the origin of plateaus \citep{2006ApJ...641L...5E, 2020MNRAS.493.3521B}
or of flares in SGRB afterglows~\citep{2022MNRAS.513..951D}. The models
explored in these studies generally require a steep jet structure in this
near-core range of angles, in apparent tension with the shallow structure
on large angles revealed in the jet of GRB170817A when fit with single
power-law structures \citep{Ghirlanda2019, gill2018}. However,
the double steep-shallow structure revealed by our simulations naturally
reconciles these two observational properties, with the steep segment
allowing for these plateaus and flare models on slightly misaligned lines
of sight and for a GRB170817A-like afterglows on significantly
misaligned ones.

Finally, we note that in the case of GRB170817, the ejecta mass inferred
from the kilonova observations is $\gtrsim 0.01 M_{\rm
  \odot}$~\citep{2017ApJ...851L..21V, 2018ApJ...856..101M, Coughlin2018a}
and the iso-equivalent energy of the core jet inferred from the afterglow
photometry and image is of $\sim 10^{52}\,{\rm erg}$, within a core
angle of $\theta_{\rm jet} \sim
6\,\rm{deg}$~\citep{Ghirlanda2019}. Assuming a jet duration of $\Delta t
= 1\,{\rm s}$ and negligible losses to SGRB prompt emission and
jet-ejecta interactions in the core jet, this would translate into a jet
luminosity of $L_{\rm jet} \sim 5 \times 10^{49}\,{\rm erg/s}$. This
places GRB170817 in the regime of our \texttt{Lj.5.49.Me.010} simulation,
where we expect significant interaction of the jet with the ejecta and a
strong effect on the high-latitude energy structure of the jet. On the
one hand, this would be compatible with the GRB170817 afterglow and its
extended slowly-increasing phase~\citep{2018ApJ...868L..11M} as due, in
our simulations, to a strong jet-ejecta interaction after launch. On the
other hand, this conclusion would cast doubts about the possibility of
using the afterglow in infer ejecta properties, since in this regime we
have shown the sensitivity to ejecta details. 

In summary, our simulations show that a great diversity of afterglow
  light-curve morphologies can arise due to the jet-ejecta interactions.
  This suggests that the afterglow of GW170817 could not be a standard
  for BNS merger afterglows, as future GW follow-up observations could
  reveal. As a result, our systematic investigation reveals that the role
  of the ejecta in shaping this afterglow provides an additional caveat
  on inference of the jet structure from the afterglow observations,
  which fortunately is less important for lower $M_{\rm ej}/L_{\rm jet}$
  ratios. Indeed, if this ratio could be determined to be small using a
  independent method (\eg based on the kilonova signal), then
  reconstructing the jet from the afterglow data will likely be more
  robust and carry less uncertainty.

\section{Conclusions} 
\label{sec:conclusion}

Using general-relativistic hydrodynamical simulations, we have examined
the propagation of an ultra-relativistic jet through an envelope of
matter ejected by a BNS merger having a variety of anisotropic
distributions of rest-mass. Overall, our results can be summarised as
follows.

\begin{itemize}

\item The geometry and degree of anisotropy of the ejecta impacts the
  propagation of the jet and influences its breakout time as well as
  the energy in the cocoon produced.

\item Contrary to expectations, the propagation of a jet encountering
  several anisotropies along its path (\ie on-axis perturbations) only
  marginally affects its breakout time, which is comparable with that in
  the absence of anisotropies.

\item The breakout time decreases as the energy of the jet is increased
  and increases as the ejected matter is more massive. This is simply
  because the jet travel-time across the ejecta will depend on its energy
  and on the matter resistance it will encounter. As a result, the most
  massive ejecta actually prevent the breakout of the weakest jet.

\item The time-averaged efficiency $\langle \epsilon \rangle$ in the
  conversion of the jet kinetic energy into the thermal energy of the
  cocoon decreases linearly with increasing energy in the jet and with
  the mass in the ejecta. This behaviour reflects the fact that more
  energetic jets are less efficient in converting the kinetic energy into
  thermal energy, especially if the ejecta are not very massive.

\item In agreement with \citet{Duffell2018}, $\langle \epsilon
  \rangle$, but also the average thermal energy $\langle E_{\rm
    th}\rangle$ and the peak internal energy $E_{\rm th, peak}$, show a
  square-root scaling with the jet energy and a linear scaling with the
  mass in the ejecta. The variance produced by the different types of
  anisotropies is marginal and most important at the largest jet
  luminosities.

\item The injection of internal energy into the matter in the cocoon by
  the propagating jet is sufficiently large to allow for the escape of
  this energy from the ejecta. The amount of energy in the cocoon that is
  able to escape $E_{\rm coc}$ grows with both the energy
  in the jet and with the mass in the ejecta. 
  
\item Given an ejecta mass and a jet propagating at a given luminosity, a
  critical efficiency of $\eta_{\rm crit} \sim {L_{\rm jet, 50}\,(t_{\rm
      jet,bo}/1\,{\rm s})}(M_{\rm ej}/{10^{-3}\,M_\odot})^{-1} \sim 0.03$ exists
  for a successful jet breakout after a time $t_{\rm jet,bo}$.
  
\item Essentially all configurations considered show a similar
  two-component behaviour for the polar structure of the jet. In
  particular, the polar profiles of the Lorentz factor consistently show
  a ``core'' with $\Gamma_\infty \sim 60-90$ within an angle of $\theta
  \approx 5\,\mathrm{deg}$. The Lorentz factor then exhibits a sharp
  decrease down to $\Gamma_\infty \lesssim 10$ for $\theta \gtrsim 7\,
  {\rm deg}$ with a power-law dependence $\Gamma_\infty(\theta) \propto
  \theta^{-\beta}$ with $\beta > 7$. The polar structure of the energy
  distribution, on the other hand, does not have a sharp fall-off but a
  much slower decay ranging from a flat distribution to power-law decays
  with index $\beta \sim 3$ up to large latitudes up to $\theta \approx
  25\,\mathrm{deg}$.

\item Within the two-components structure of the jet, the core is
  essentially unaffected by the anisotropies in the ejecta, while the
  shallow segment displays the greatest variance, which is stronger for
  lower ratios of $L_{\mathrm{jet}} / M_{\mathrm{ej}}$. In the case of
  the marginally successful jet, on the other hand, the
  two-components structure is essentially not present and the polar
  distributions of energy and Lorentz factor are far more irregular.
  
\item Different anisotropies impact the variance of the afterglow
  light-curves as a function of the jet luminosity and ejected mass. In
  particular, the variance is reduced as the injected energy in the jet
  is increased $L_{\rm jet}$, but it increases with the mass in the
  ejecta $M_{\rm ej}$. As a result, the impact of the anisotropies in the
  ejecta effectively increases with the ratio $M_{\rm ej}/L_{\rm jet}$.

\item The anisotropies in the afterglow observations are most
  dominant in the first tens of days, for off-axis observations of
  systems with generally high ejecta mass. In the case of BNS mergers,
  strong candidates for detection of these effects, are the systems with
  low mass ratio $q := M_2/M_1$, where $M_1$ is the mass of the primary
  and $M_2$ that of the secondary. The reason for this is that these
  systems generally produce larger amounts of dynamically ejected matter.

\item Double-peaks in the light-curves can also arise from the polar
  structure of the energy distribution of the jet, provided it has two
  portions, with a steep inner core and a shallow exterior with a local
  maximum at large latitudes. This represents a further illustration of
  the degeneracies that can be present when in inferring jet structures
  from afterglow light-curves.
  
\item The presence of an extended shallow segment in the jet structure
  for systems with moderate $L_{\rm jet}/M_{\rm ej}$ ratios results in
  long-lasting phases of slow brightening of the afterglow as the
  emission is dominated by material progressively ranging up the shallow
  segment (\cf GRB170817A).

\item In the case of GRB170817A, assuming a jet duration of $\Delta t =
  1\,{\rm s}$ and negligible losses to GRB prompt emission and jet-ejecta
  interactions in the core jet, leads to a jet luminosity of $L_{\rm jet}
  \sim 5 \times 10^{49}\,{\rm erg/s}$. This is in the regime where we
  expect significant interaction of the jet with the ejecta and a strong
  effect on the high-latitude energy structure of the jet.
  
\end{itemize}

While the present work arguably represents the most extended and
systematic investigation of the interaction of a relativistic jet with
the ejecta from a BNS merger with non-smooth features in the density
distribution, it can be improved in at least three different
directions. First, by including the presence of a magnetic field of
different topologies and strengths~\citep[][]{nathanail2020b,
  2021MNRAS.502.1843N}, which could influence the propagation of the jet
across the ejecta and that, in the presence of turbulence-driven
reconnection, could lead to an increase of the internal energy of the
cocoon. Second, by extending to three the number of spatial dimensions,
which would improve the description of the dynamics of the jet head and
remove some of the artefacts discussed here. Finally, by moving away from
the simple equation of state considered in our work and by considering
instead the same equations of state describing the neutron-star matter in
full numerical-relativity calculations. This final step, which is less
trivial than it may appear given the complexity of the equations of state
employed in BNS mergers, will finally allow for a consistent and faithful
import of realistic ejecta distributions from merger simulations.

As a concluding remark, we note how future gravitational-wave observing
runs will see more afterglows of jets with significantly off-axis lines
of sight, both with current facilities
\citep[e.g.,][]{2019A&A...631A..39D, 2022ApJ...937...79C} and with
next-generation instruments \citep{2022A&A...665A..97R}, thus allowing
for deeper insights into the jet structure. In view of our results, the
combination of this data with kilonovae signals \citep[see,
  \eg][]{2021A&A...651A..83M} could provide a very powerful tool to
understand the properties of the sources. Indeed, if dim kilonova signals
were to be more often associated with rapidly-increasing afterglows, they
could provide the signature of a small ejecta mass weakly imprinting the
incipient jet, which therefore retained its steep initial structure as
described here. In this case, the incipient jet structures could be
considered a faithful probe of the jet-launching physics at the scales of
the central engine. If, on the contrary, dim kilonovae were to be most
regularly associated with slowly-increasing afterglows, they could hint
to the fact that the associated shallow jet structures are acquired
directly at jet launch. Finally, the inclusion of gravitational-wave
constraints on the progenitor binaries can help bridge the gap between
the kilonova and afterglow jet signals with the theoretical predictions 
obtained by numerical simulations on the ability of magnetised
merging BNSs to launch relativistic jets~\citep[see,
  \eg][]{rezzolla:2011, 2013PhRvD..87b4001H, Palenzuela2015, Ruiz2016}.

\section*{Acknowledgements}

Support comes from the ERC Advanced Grant: ``JETSET: Launching,
propagation and emission of relativistic jets from binary mergers and
across mass scales'' (Grant No. 884631) and from the State of Hesse
within the Research Cluster ELEMENTS (Project ID 500/10.006). LR
acknowledges the Walter Greiner Gesellschaft zur F\"orderung der
physikalischen Grundlagenforschung e.V. through the Carl W. Fueck
Laureatus Chair. The simulations were performed on SuperMUC at LRZ in
Garching, on the Calea cluster at the ITP in Frankfurt, and on the HPE
Apollo Hawk at the High Performance Computing Center Stuttgart (HLRS)
under the grant numbers BBHDISKS and BNSMIC.

\section*{Data Availability}
The data underlying this article will be shared on reasonable request to
the corresponding author.


\bibliographystyle{mnras}
\bibliography{pert}

\newpage
\appendix

\section{Numerical consistency and robustness of the results}
\label{sec:conv}

\let\origaddcontentsline\addcontentsline
\def\addcontentsline#1#2#3{\origaddcontentsline{#1}{#2}{#3}\let\addcontentsline\origaddcontentsline}

\begin{figure*}
\centering
\includegraphics[width=0.9\textwidth]{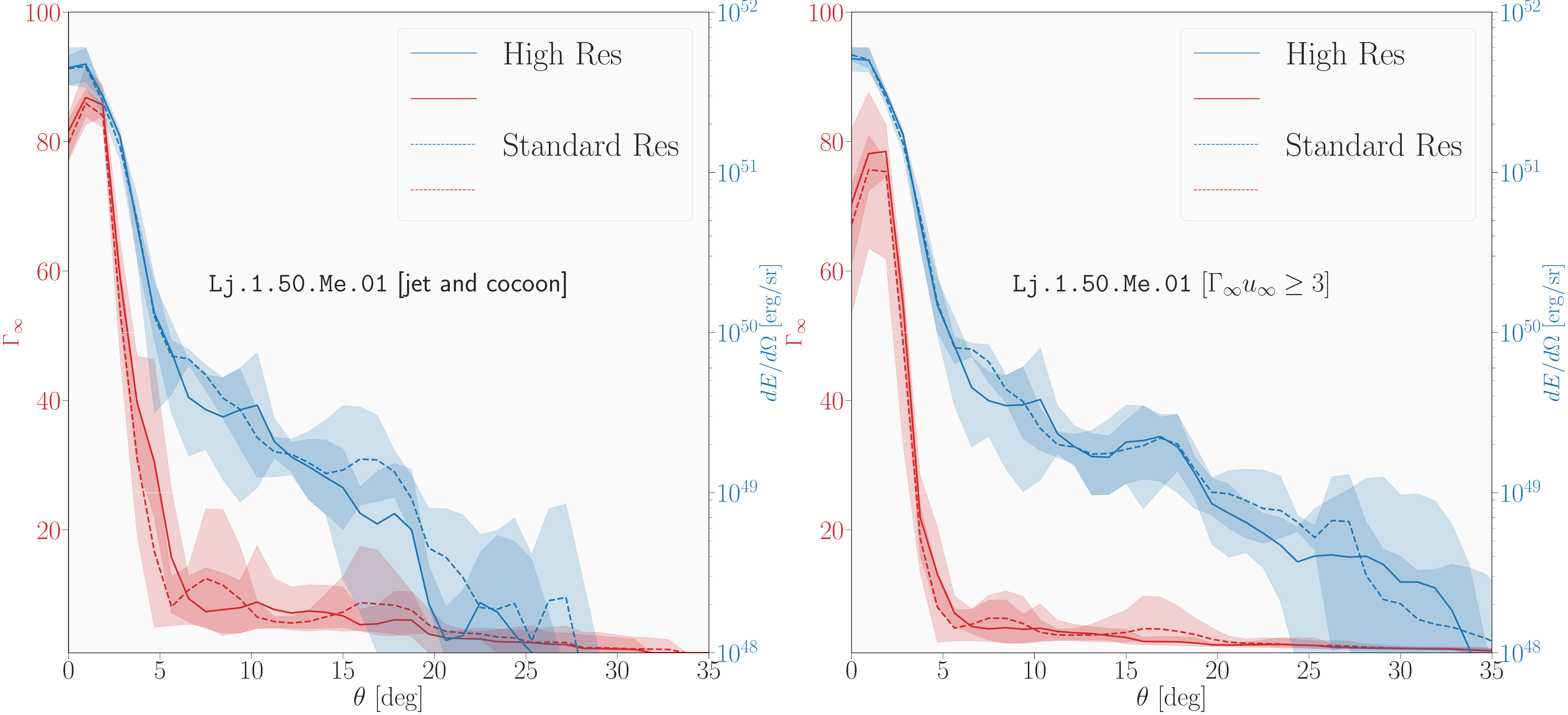} 
\caption{The same as in Fig.~\ref{fig:prof} for model
  \texttt{Lj.1.50.Me.010} at $t=2.25\,\mathrm{s}$ after jet
  launching. The solid and dashed lines refer to simulations performed
  with a numerical resolution that is either ``high'' or ``standard'',
  respectively. Clearly, the qualitative features of the polar structures
  are robust and no new features are introduced when increasing the
  resolution.}
\label{profile_conv}
\end{figure*}

\let\addcontentsline\origaddcontentsline

In this Appendix, we study the consistency and the robustness of our
results to the choice of resolution for our numerical simulations. We
focus on the main result extracted from our simulation domain: the energy
and Lorentz-factor structures of the relativistic jets after breakout
from the ejecta.

We proceed with the method described in Sec.~\ref{sec:agmeth} considering
two different numerical resolutions: the first, referred to as ``standard
resolution'', is as provided in Sec.~\ref{sec:bhac} with a mesh of $N_r
\times N_{\theta} = 1120 \times 576$ cells and three refinement levels;
the second resolution, dubbed ``high resolution'', is 1.5 times finer
with a base grid of $N_r \times N_{\theta} = 1680 \times 864$ and three
adaptive refinement levels.

In Fig.~\ref{profile_conv}, we report the results for the
\texttt{Lj.1.50.Me.010} system using the two resolution levels,
presenting the energy ${\rm d}E_{\rm jet}/{\rm d}\Omega(\theta)$ and
Lorentz-factor, $\Gamma_\infty(\theta)$ structures in analogy with
Fig.~\ref{fig:prof} of the main text. As in the main text, we report the
structured obtained by applying two different cuts in the Lorentz-factor,
aiming to capture the jet and cocoon material, or only the high-velocity
jet.  We find that, within the variance induced by the three choices of
the density profiles of the eject, the results are consistent for the two
resolutions. This is the case for both the choices of velocity cuts, with
the highest fidelity between the two resolutions found for the
high-velocity structures.

In particular, the presence of a large energy
reservoir at high-latitudes is a robust feature, and the energy level and
angular extent of this reservoir is consistent between the runs with
different resolution. The run presented in Fig.~\ref{profile_conv} has
the largest ejecta mass among our runs, such that the effect on the jet
is expected to be among the largest. Since this behaviour is properly
captured and exhibits consistency between the two resolutions, we
conclude that the results of the other runs are also robust and can be
used to derive reliably the conclusions drawn in the main text.

\label{lastpage}
\end{document}